\newtheorem{Theorem}{Theorem}
\documentclass[12pt,journal,onecolumn]{IEEEtran}
\usepackage{times,amsmath,amssymb,dsfont,graphicx,float, enumerate}

\newcommand{\lsN}{{\log\sqrt{N}}}
\newcommand{\ep}{\varepsilon}
\ifCLASSINFOpdf
\else
\fi
\IEEEoverridecommandlockouts

\begin{document}

\title{Error Probability Bounds for \\Balanced Binary Relay Trees}

\author{Zhenliang~Zhang,~\IEEEmembership{Student~Member,~IEEE,}
        Ali~Pezeshki,~\IEEEmembership{Member,~IEEE,}
        William~Moran,~\IEEEmembership{Member,~IEEE,}
        Stephen~D.~Howard,~\IEEEmembership{Member,~IEEE,}
        and~Edwin~K.~P.~Chong,~\IEEEmembership{Fellow,~IEEE}
\thanks{This work was supported in part by AFOSR under Contract FA9550-09-1-0518, and by NSF under Grants ECCS-0700559, CCF-0916314, and CCF-1018472.}
\thanks{Z. Zhang, A. Pezeshki, and E. K. P. Chong are with the Department
of Electrical and Computer Engineering, Colorado State University, Fort Collins, CO, 80523-1373, USA
(e-mail: Zhenliang.Zhang@colostate.edu; Ali.Pezeshki@colostate.edu; Edwin.Chong@colostate.edu)
}
\thanks{W. Moran is with the Department of Electrical and Electronic Engineering, The University of Melbourne, Melbourne, Vic., 3010, Australia (e-mail: b.moran@ee.unimelb.edu.au)
}

\thanks{S. D. Howard is with the Defence Science and Technology Organization, P.O.
Box 1500, Edinburgh, SA, 5111, Australia (e-mail: sdhoward@unimelb.edu.au)}
}

\maketitle

\begin{abstract}
We study the detection error probability associated with
a balanced binary relay tree, where the leaves of the tree correspond
to  $N$ identical and independent detectors.
The root of the tree represents a fusion center that makes the overall detection decision.
Each of the other nodes in the tree are relay nodes that combine
two binary messages to form a single output binary message. In this
way, the information from the detectors is aggregated into
the fusion center via the intermediate relay nodes.
In this context, we describe the evolution of Type I and Type II error probabilities
of the binary data as it propagates from the leaves  towards the root. Tight
upper and lower bounds for the total error probability at the fusion
center as functions of $N$ are derived. These characterize how fast the total error
probability converges to $0$ with respect to $N$, even if the individual sensors have error probabilities that
converge to $1/2$.
\end{abstract}


\section{Introduction}

Consider a hypothesis testing problem under two scenarios:
\emph{Centralized} and \emph{decentralized}. Under the centralized
network scenario, all sensors send their raw measurements to the
fusion center which makes a decision based on these
measurements. In the decentralized network introduced
in~\cite{Tenney}, sensors send summaries of their measurements and
observations to the fusion center. The fusion center then makes a
decision. In a decentralized network, information
is summarized into smaller messages.  Evidently, the decentralized network cannot perform
better than the centralized network. It gains  because of its  limited use
of  resources and bandwidth; through  transmission of  summarized information
it is more practical and efficient.

The decentralized network in \cite{Tenney} involves a parallel
architecture,  known as a \emph{star} architecture
\cite{Tenney}--\nocite{Chair,Cham,dec,Tsi,Tsi1,Warren,Vis,Poor,Sah,chen1,Liu,chen,Kas}\cite{Chong},\cite{BOOK},
in which all sensors directly connect to the fusion center.  A typical
result is that under the assumption of (conditionally) independence
 of the sensor observations, the decay rate of the error probability in a
parallel network is exponential \cite{Tsi1}.

Several different sensor topologies have been studied under the
assumption of conditional independence. The first configuration for
such a fusion network
considered was the tandem network
\cite{Tang}--\nocite{Tum,tandem,athans}\cite{Venu},\cite{BOOK}. In
such a  network, each non-leaf node combines the information from its
own sensor with the message it has received from the node at one level
down, which is then transmitted to the node at the next level up. The
decay rate of the error probability in this case is sub-exponential
\cite{Venu}.  As the number of sensors $N$ goes to
infinity, the error probability is $\Omega(e^{-cN^d})$ for some
positive constant $c$ and for all $d\in (1/2,
1)$\space\cite{tandem}. This sensor network represents a situation
where  the length of the network is the longest
possible among all networks with $N$ leaf nodes.

The asymptotic performance of single-rooted tree networks with bounded
height is discussed in \cite{Tang1}--\nocite{Nolte,tree1,tree2,tree3,Pete,Alh,Will}\cite{Lin}\cite{BOOK}.
Even though error probabilities in the parallel configuration decrease
exponentially, in a practical implementation,
the resources  consumed in having each sensor transmit
directly to the fusion center might be regarded as excessive.  Energy consumption
can be reduced by setting up a directed tree, rooted at the fusion
center.  In this tree structure, measurements are summarized by leaf
sensor nodes and sent to their parent nodes, each of which fuses all
the messages it receives with its own measurement (if any) and then
forwards the new message to its parent node at the next level.  This
process takes place throughout the tree culminating in  the fusion center,
where a final decision is made. For a bounded-height tree, the error
exponent is as good as that of the parallel configuration under
certain conditions. For example, for a bounded-height tree network
with $\lim _{\tau_N\rightarrow \infty} \ell_N/\tau_N=1$, where $\tau_N$ denotes the total number of nodes and $\ell_N$ denotes the number of leaf nodes, the
optimum error exponent is the same as that of the parallel
configuration \cite{tree1}.

The variation of detection performance with increasing tree height  is still
largely unexplored. If only the leaf nodes have sensors  making observations, and
all other nodes simply fuse the messages received and forward the new
messages to their parents, the tree network is known as a \emph{relay tree}.
The balanced binary relay tree has been addressed in
\cite{Gubner}. Evidently, the height of this tree is $\log_2 N$. In \cite{Gubner}, it is
assumed that the leaf nodes are independent
sensors with identical Type I error probability ($\alpha_0$) and
identical Type II error probability ($\beta_0$).   It is shown there
 that if sensor error probabilities satisfy the condition
$\alpha_0+\beta_0<1$, then both the Type I and Type II error probabilities at
the fusion center both converge to 0 as the $N$ goes to infinity.  If
$\alpha_0+\beta_0>1$, then both Type I and Type II error probabilities
converge to 1.

We consider the same tree configuration in this paper and describe the
precise evolution of Type I and Type II error probabilities in this case. In
addition, we provide upper and lower bounds for the total error
probability at the fusion center as functions of $N$. These
characterize the decay rate of the total error probability. We also
show that the total error probability converges to $0$ under certain
condition even if the sensors are asymptotically crummy, that is,
$\alpha_0+\beta_0 \rightarrow 1$.

The rest of the paper is organized as follows. In Section II, we
formulate the decentralized detection problem in the setting of balanced binary relay
trees. In Section III, we discuss the evolution of Type I and Type II
error probabilities. In Section IV, we derive upper and lower bounds
for the total error probability at the fusion center as functions of
$N$.  In Section V, we discuss some corollaries focusing on the
asymptotic regime as $N\rightarrow \infty$. Finally, Section VI
contains the concluding remarks.

\section{Problem Formulation}
We consider the problem of binary hypothesis testing between $H_0$ and
$H_1$ in a balanced binary relay tree. Leaf nodes are sensors
undertaking initial and independent detections of the same event in a
scene.  These measurements are summarized into binary messages and
forwarded to nodes at the next level. Each non-leaf node with the
exception of the root, the fusion center, is a relay node, which fuses
two binary messages into one new binary message and forwards the new
binary message to its parent node. This process takes place at each
intermediate node culminating in the fusion center,  at which the
final decision is made based on the information received.

\begin{figure}[htbp]
\centering
\includegraphics[width=4.5in]{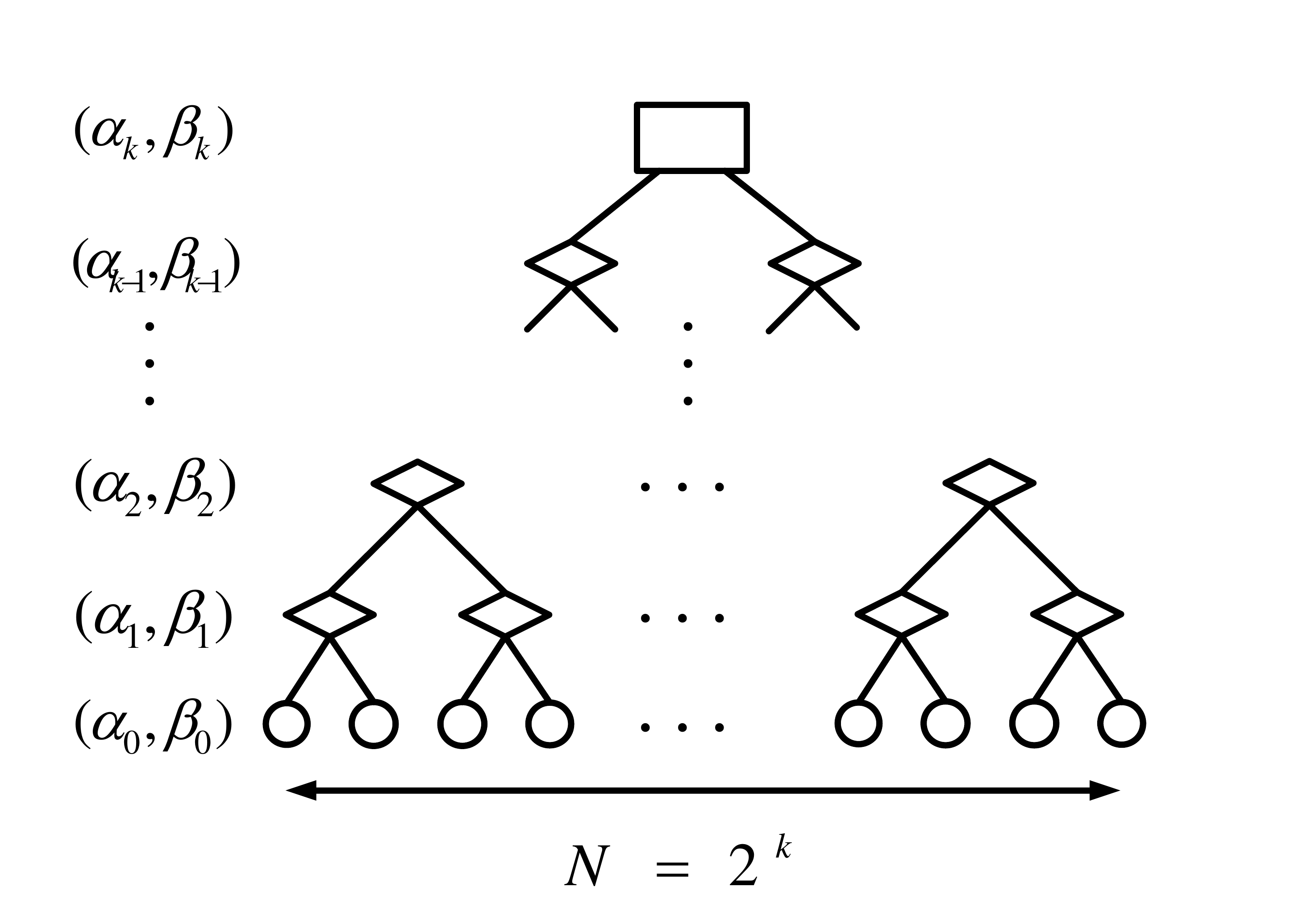}
\caption{A balanced binary relay tree with height $k$. Circles represent sensors making measurements. Diamonds represent relay nodes which fuse binary messages. The rectangle at the root represents the fusion center making an overall decision.}
\label{fig:tree}
\end{figure}

In this configuration, as shown in Fig. \ref{fig:tree}, the closest
sensor to the fusion center is as far as it could be, in terms of the
number of arcs in the path to the root.  In this sense, this
configuration is the worst case among all $N$ sensor relay trees.
Moreover, in contrast to the configuration in \cite{tree1} discussed earlier, in our balanced binary tree we have $\lim _{\tau_N\rightarrow \infty} \ell_N/\tau_N = 1/2$ (as opposed to $1$ in \cite{tree1}). Hence, the number of times that information is aggregated here is
essentially as large as the number of measurements (cf., \cite{tree1}, in which the number of measurements dominates the number of fusions).

We assume
that all sensors are independent given each hypothesis, and that  all
sensors have identical Type I error probability $\alpha_0$ and
identical Type II error probability $\beta_0$.  The likelihood-ratio test \cite{VT} is applied
as the fusion rule at the intermediate relay nodes and at the fusion
center. We are interested in following questions:

\begin{itemize}
\item What are these Type
  I and Type II error probabilities as functions of $N$?
\item Will they
    converge to $0$ at the fusion center?
\item If yes, how fast will they converge with respect to $N$?
\end{itemize}

Fusion at a single node receiving information from the
 two immediate child nodes where  these  have
identical Type I error probabilities $\alpha$ and identical Type II
error probabilities $\beta$. provides a detection with  Type I and Type II error probabilities
 denoted by $(\alpha', \beta')$, and given by
\cite{Gubner}:
\begin{equation}
(\alpha', \beta')=f(\alpha, \beta):=\left\{\begin{array}{c}
(1-(1-\alpha)^2, \beta^2), \quad \alpha\leq\beta,\\
 \\
(\alpha^2, 1-(1-\beta)^2), \quad \alpha>\beta.\end{array}\right.
\label{equ:rec}
\end{equation}

Evidently, as all sensors have the same error probability pair
$(\alpha_0,\beta_0)$,  all relay nodes at level $1$ will have the
same error probability pair $(\alpha_1,\beta_1)=f(\alpha_0,\beta_0)$,
and by recursion,
\begin{equation}
(\alpha_{k+1}, \beta_{k+1})=f(\alpha_k, \beta_k), \quad \quad k=0,1,\ldots,
\label{equ:rel}
\end{equation}
where $(\alpha_k,\beta_k)$ is the error probability pair of nodes at the $k$th level of the tree.

The recursive relation (\ref{equ:rel}) allows us to consider the pair
of Type I and II error probabilities as a discrete dynamic system.  In
\cite{Gubner}, which focuses on the convergence issues for
the total error probability, convergence was proved
using Lyapunov methods.  The analysis of the precise evolution
of the sequence $\{(\alpha_k, \beta_k)\}$ and the total error
probability decay rate remain open.  In this paper, we will establish
upper and lower bounds for the total error probability and deduce the
precise decay rate of the total error probability.

\begin{figure}[htbp]
\centering
\includegraphics[width=3in]{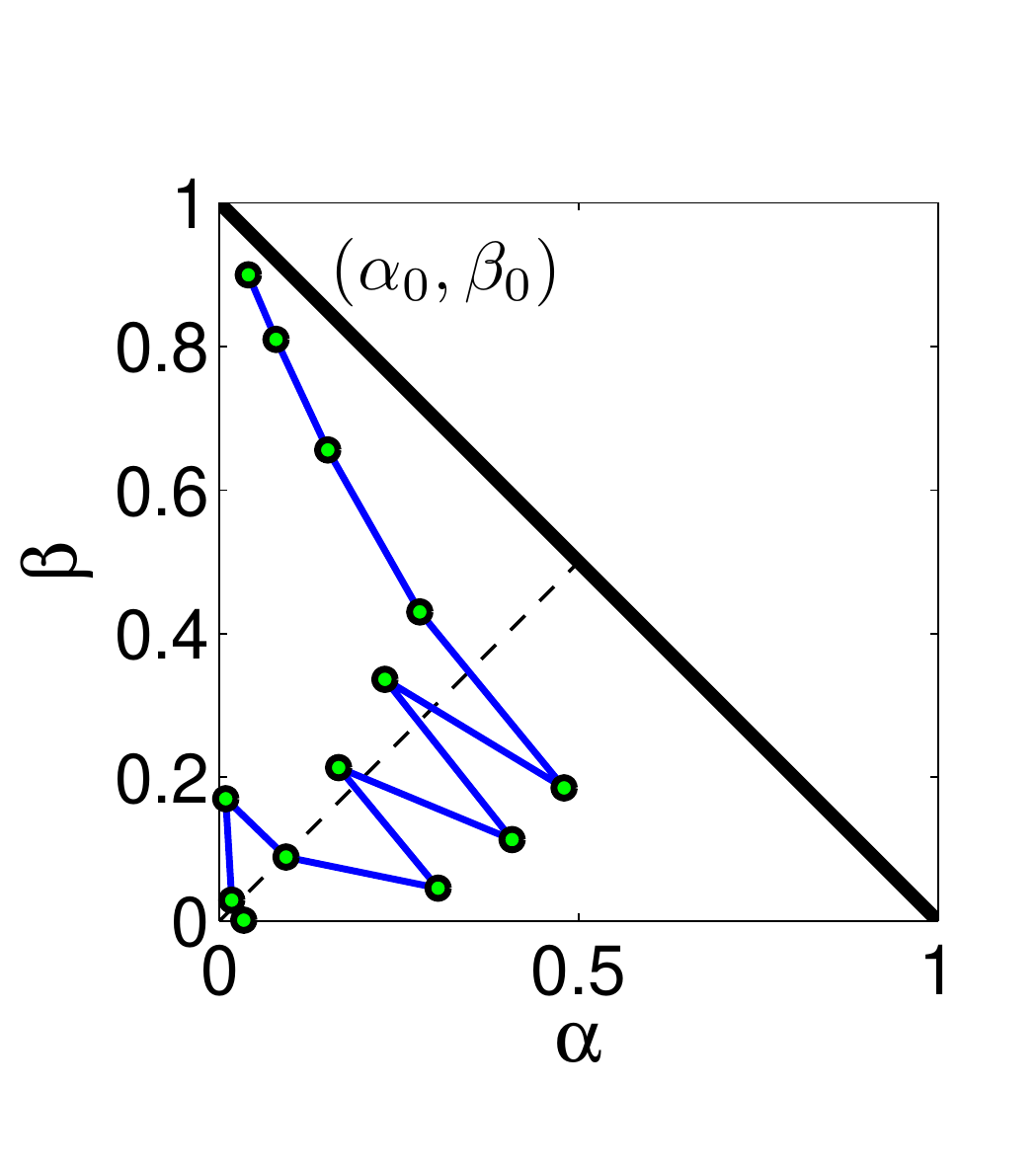}
\caption{A trajectory of the sequence $\{(\alpha_k,\beta_k)\}$ in the $(\alpha,\beta)$ plane.}
\label{fig:plane}
\end{figure}

To illustrate the ideas, consider first a  single trajectory
for the dynamic system given by equation \eqref{equ:rec}, and starting at the
initial state $(\alpha_0, \beta_0)$.  This trajectory is
shown in Fig. \ref{fig:plane}.    It
exhibits different behaviors depending on its distance from the
$\beta=\alpha$ line. The trajectory approaches $\beta=\alpha$ very
fast initially,  but when  $(\alpha_k,\beta_k)$  approaches within a
certain neighborhood of the line
$\beta=\alpha$, the next pair $(\alpha_{k+1},\beta_{k+1})$ will appear
on the other side of that line.  In the
next section, we will establish theorems that characterize the precise
step-by-step behavior of the dynamic system (\ref{equ:rel}).

\section{Evolution of error probabilities}

The relation (\ref{equ:rec}) is symmetric about both of the lines
$\alpha+\beta = 1$ and $\beta=\alpha$. Thus, it suffices to study the
evolution of the dynamic system only in the region bounded by
$\alpha+\beta < 1$ and $\beta \geq \alpha$. Let
$\mathcal{U}:=\{(\alpha, \beta)\geq 0|\alpha+\beta<1 \text{\space and
  \space} \beta\geq\alpha\}$ be this triangular region. Similarly,
define the complementary triangular region $\mathcal{L}:=\{(\alpha,
\beta)\geq 0|\alpha+\beta<1 \text{\space and \space} \beta<\alpha\}$.

\begin{figure}[tp]
\begin{center}
\begin{tabular}{cc}
\includegraphics[width=3in]{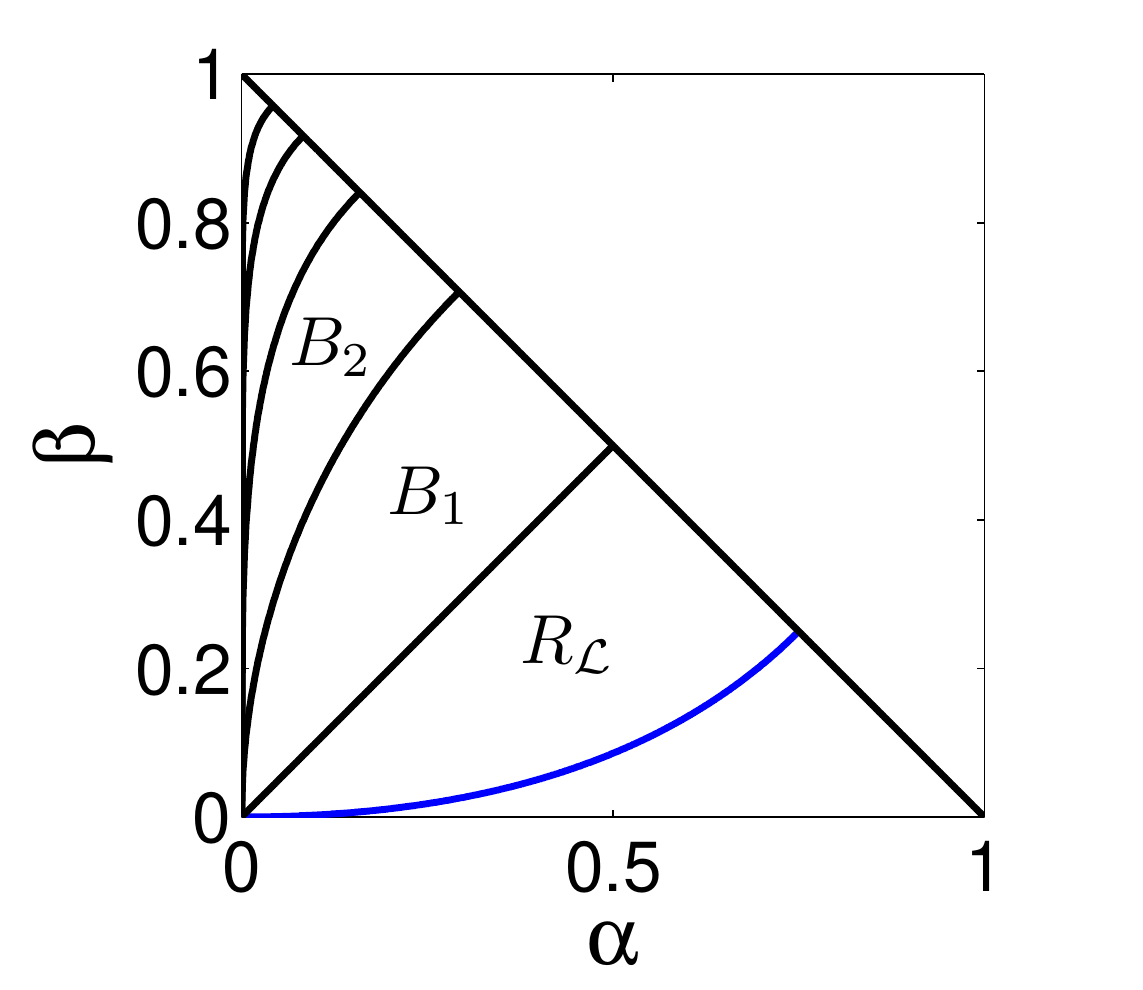} & \includegraphics[width=3in]{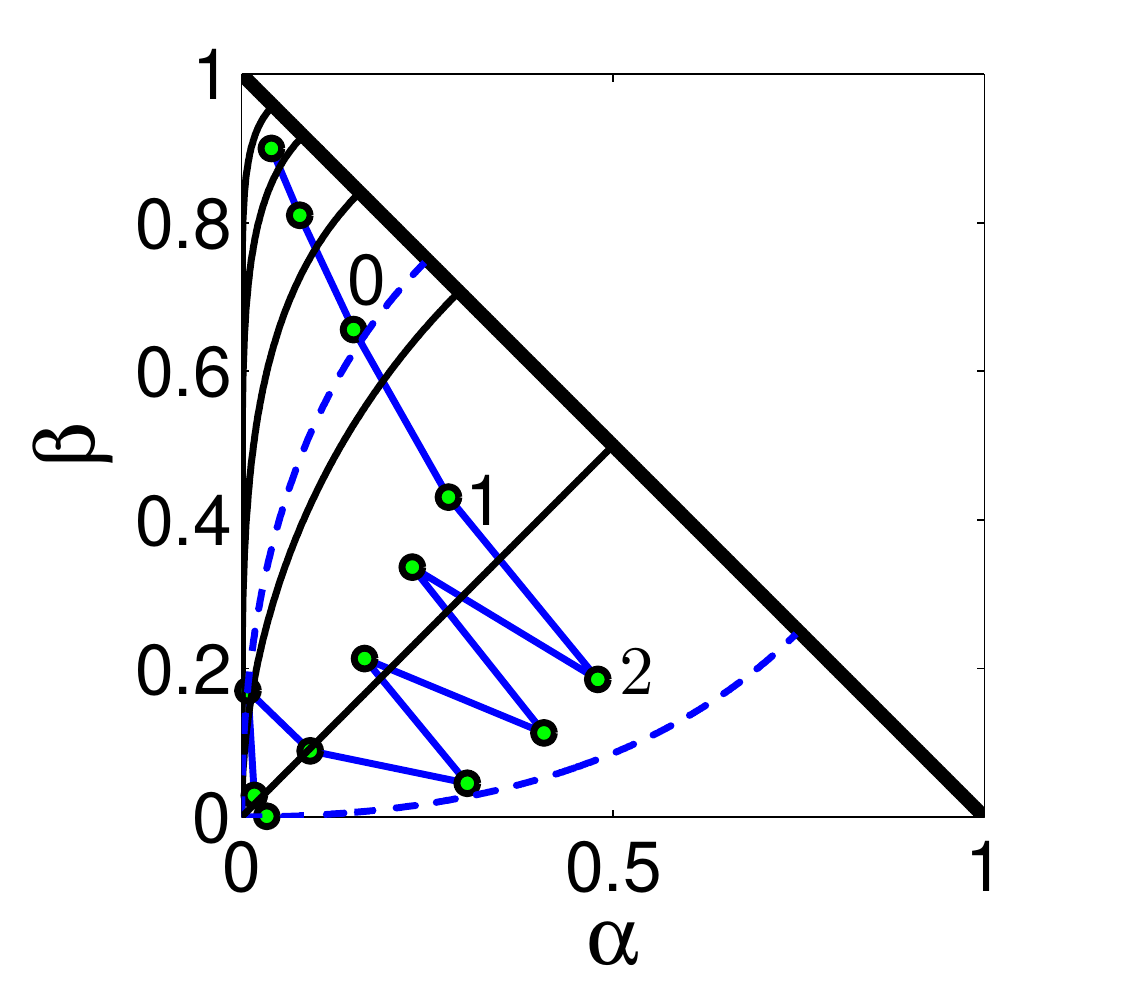}\\
(a) & (b)\\
\end{tabular}
\end{center}
\caption{(a) Regions $B_1$, $B_2$, and $R_\mathcal{L}$ in the $(\alpha, \beta)$ plane. (b) The trajectory in Fig. \ref{fig:plane} superimposed on (a), where solid lines represent boundaries of $B_m$ and dashed lines represent boundaries of $R$.}\label{fig:b}
\end{figure}

Let $B_1:=\{(\alpha, \beta)\in \mathcal{U}|(1-\alpha)^2+\beta^2\leq
1\}$.  If $(\alpha_k,\beta_k)\in B_1$, then the next pair
$(\alpha_{k+1},\beta_{k+1})=f(\alpha_k,\beta_k)$ crosses  the line
$\beta=\alpha$  to the opposite side from $(\alpha_k,\beta_k)$. More precisely, if
$(\alpha_k,\beta_k)\in \mathcal{U}$, then $(\alpha_k,\beta_k)\in B_1$
if and only if $(\alpha_{k+1},\beta_{k+1})=f(\alpha_k,\beta_k)\in
\mathcal{L}$.  In other words, $B_1$ is the \emph{inverse image} of
$\mathcal{L}$ under $f$ in $\mathcal{U}$. The set $B_1$ is shown
in Fig.~\ref{fig:b}(a). Fig.~\ref{fig:b}(b) illustrates this behavior
of the trajectory for the example in Fig.~\ref{fig:plane}. For
instance, as shown in Fig.~\ref{fig:b}(b), if the state is at point 1
in $B_1$, then it jumps to the next state point 2, on the other side
of $\beta=\alpha$.

Let $B_2:=\{(\alpha,
\beta)\in \mathcal{U}| (1-\alpha)^2+\beta^2\geq 1 \text{\space and
  \space} (1-\alpha)^4+\beta^4\leq 1\}$. It is easy to show that if
$(\alpha_k,\beta_k)\in \mathcal{U}$, then $(\alpha_k,\beta_k)\in B_2$
if and only if $(\alpha_{k+1},\beta_{k+1})=f(\alpha_k,\beta_k)\in
B_1$. In other words, $B_2$ is the inverse image of $B_1$ in
$\mathcal{U}$ under $f$. The regions and the behavior of $f$ is
illustrated in the  movement from  $0$ to
point $1$ in Fig.~\ref{fig:b}(b). The set $B_2$ is identified in
Fig.~\ref{fig:b}(a),  lying directly above $B_1$.

Now for an integer  $m>1$,  recursively define $B_m$ to be  the inverse image of $B_{m-1}$ under $f$, denoted
by $B_m$. It is easy to see that
$B_m:=\{(\alpha, \beta)\in \mathcal{U}|
(1-\alpha)^{2^{(m-1)}}+\beta^{2^{(m-1)}}\geq1 \text{\space and \space}
(1-\alpha)^{2^m}+\beta^{2^m}\leq1\}$.  Notice that
$\mathcal{U}=\bigcup_{m=1}^\infty B_m$. Hence, for any
$(\alpha_0,\beta_0)\in \mathcal{U}$, there exists $m$ such that
$(\alpha_0, \beta_0)\in B_m$.  This gives a complete description of
how the dynamics of the system
behaves in the upper triangular region $\mathcal{U}$. For instance, if
the initial pair $(\alpha_0,\beta_0)$ lies in $B_m$, then the system
evolves in the order
\[
B_m\rightarrow B_{m-1}  \rightarrow \cdots \rightarrow B_{2}\rightarrow B_1.
\]
Therefore, the system will enter $B_1$ after $m-1$ levels of fusion,
i.e., $(\alpha_{m-1},\beta_{m-1})\in B_1$.

As the next stage,  we consider the behavior of the system after it
enters $B_1$.
The \emph{image} of
$B_1$ under $f$, denoted by $R_\mathcal{L}$, is (see Fig. \ref{fig:b}
(a))
\begin{equation}
  \label{eq:1}
R_\mathcal{L}:=\{(\alpha, \beta)\in
\mathcal{L}|\sqrt{1-\alpha}+\sqrt{\beta}\geq1\}
\end{equation}

\begin{figure}[bp]
\centering
\includegraphics[width=3.5in]{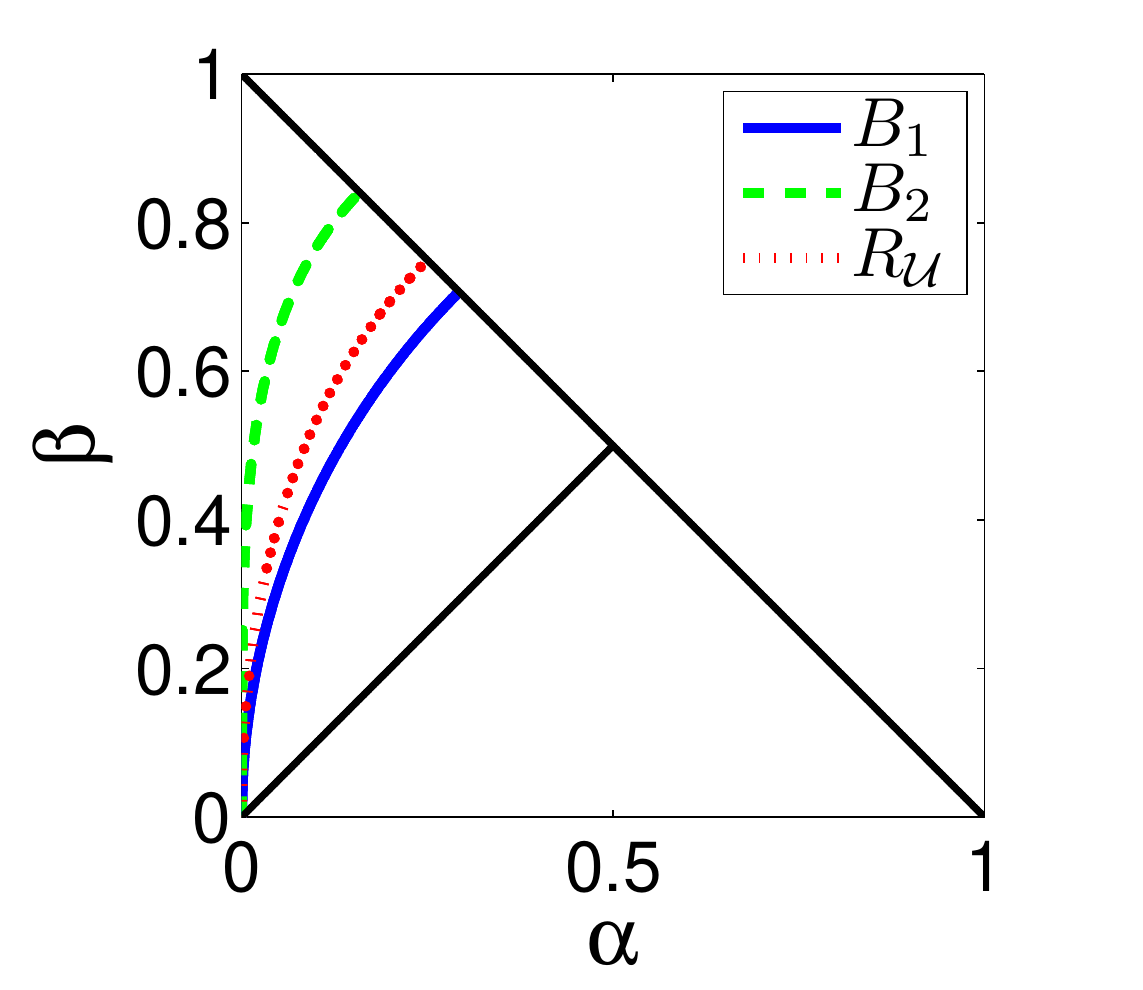}
\caption{Upper boundaries for $B_1, B_2, \text{ and }R_\mathcal{U}$.}
\label{fig:ru}
\end{figure}

We can define the reflection of $B_m$ about the line $\beta=\alpha$ in
the similar way for all $m$.
Similarly, we define the reflection of
$R_\mathcal{L}$ about the line $\beta=\alpha$ to be
$R_\mathcal{U}:=\{(\alpha,\beta)\in
\mathcal{U}|\sqrt{1-\beta}+\sqrt{\alpha} \geq 1\}$. We  denote the
region $R_\mathcal{U} \cup R_\mathcal{L}$ by $R$.  Below
$R$ is shown to be an \emph{invariant region} in the sense that once the system
enters $R$, it stays there. For example, as shown in
Fig.~\ref{fig:b}(b), the system after point $1$ stays inside $R$.

\emph{Proposition 1}: If $(\alpha_{k_0},\beta_{k_0})\in R$ for some $k_0$, then $(\alpha_k,\beta_k)\in R$ for all $k\geq k_0$.
\begin{IEEEproof}
  First we show that $B_1\subset R_\mathcal{U} \subset B_1\cup B_2$.

Notice that $B_1$, $R_\mathcal{U}$, and $B_1\cup B_2$ share the same lower boundary $\beta=\alpha$.
It suffices to show that the upper boundary of $R_\mathcal{U}$ lies
between
the upper boundary of $B_2$ and that of $B_1$ (see Fig.~\ref{fig:ru}).

First, we show that the upper boundary of
$R_\mathcal{U}$ lies above the upper boundary of $B_1$. We have
\begin{eqnarray*}
&& 1-(1-\sqrt{\alpha})^2 \geq \sqrt{1-(1-\alpha)^2}\\
&\Longleftrightarrow& 2\sqrt{\alpha}-\alpha \geq \sqrt{2\alpha-\alpha^2}\\
&\Longleftrightarrow& \alpha^2+\alpha-2\alpha^{\frac{3}{2}} \geq 0,
\end{eqnarray*}
which holds for all $\alpha$ in $[0,1)$. Thus, $B_1\subset R_\mathcal{U}$.

Now we prove that the upper boundary of $R_\mathcal{U}$ lies below that of $B_2$. We have
\begin{eqnarray*}
&& (1-(1-\alpha)^4)^{\frac{1}{4}} \geq 1-(1-\sqrt{\alpha})^2\\
&\Longleftrightarrow& 1-(1-\alpha)^4 \geq (2\sqrt{\alpha}-\alpha)^4 \\
&\Longleftrightarrow&-2(\sqrt{\alpha}-1)^2\alpha(-\alpha^{\frac{3}{2}}+\alpha(\sqrt{\alpha}-1)+4\sqrt{\alpha}(\sqrt{\alpha}-1)+\alpha-2) \geq 0,
\end{eqnarray*}
which holds for all $\alpha$ in $[0,1)$ as well. Hence, $R_\mathcal{U}\subset B_1 \cup B_2$.

Without lost of generality, we assume that $(\alpha_{k_0},
\beta_{k_0})\in R_\mathcal{U}$. That means $(\alpha_{k_0},
\beta_{k_0})\in B_1$ or $(\alpha_{k_0}, \beta_{k_0})\in B_2\cap
R_\mathcal{U}$. If $(\alpha_{k_0}, \beta_{k_0})\in B_1$, then the next
pair $(\alpha_{k_0+1}, \beta_{k_0+1})$ is in $R_\mathcal{L}$. If
$(\alpha_{k_0}, \beta_{k_0})\in B_2\cap R_\mathcal{U}$, then
$(\alpha_{k_0+1}, \beta_{k_0+1})\in B_1\subset R_\mathcal{U}$ and
$(\alpha_{k_0+2}, \beta_{k_0+2})\in R_\mathcal{L}$.  By symmetry
considerations, it follows  that the system stays inside $R$ for all $k\geq k_0$.

\end{IEEEproof}

So far we have studied the precise evolution of the sequence
$\{(\alpha_k,\beta_k)\}$ in the $(\alpha, \beta)$ plane. In the next
section, we will consider the step-wise reduction in the total error
probability and deduce upper and lower bounds for it.

\section{Error probability bounds}
The total error probability for a node with $(\alpha_k, \beta_k)$ is
$(\alpha_k+\beta_k)/2$ because of the equal prior assumption. Let
$L_k=\alpha_k+\beta_k$, namely,  twice the total error probability.
Analysis of the total error probability will result from consideration
of the sequence $\{L_k\}$. In fact, we will derive bounds on
$\log L_k^{-1}$, whose growth rate is related to the rate of converge
of $L_k$ to $0$. (Throughout this paper, $\log$ stands for the binary
logarithm.)  We divide our analysis into two parts:

\begin{enumerate}[I]
\item  We will study the shrinkage of the total error probability as the
system propagates from $B_m$ to $B_1$;
\item  We will study the shrinkage of the total error probability
  after the system enters $B_1$.
\end{enumerate}

\subsection{Case I: Error probability analysis as the system propagates from $B_m$ to $B_1$}
Suppose that the initial state $(\alpha_0, \beta_0)$ lies in $B_m$,
where $m$ is a positive integer and $m\neq1$. From the previous
analysis,  $(\alpha_{m-1}, \beta_{m-1}) \in B_1$. In this section,
we study the rate of reduction of  the total error probability as the system
propagates from $B_m$ to $B_1$.

\emph{Proposition 2}:
Suppose that $(\alpha_k, \beta_k) \in B_m$, where $m$ is a positive integer and $m\neq1$. Then,

\[
1\leq \frac{L_{k+1}}{L_{k}^2} \leq2.
\]

\begin{figure}[htbp]
\centering
\includegraphics[width=3in]{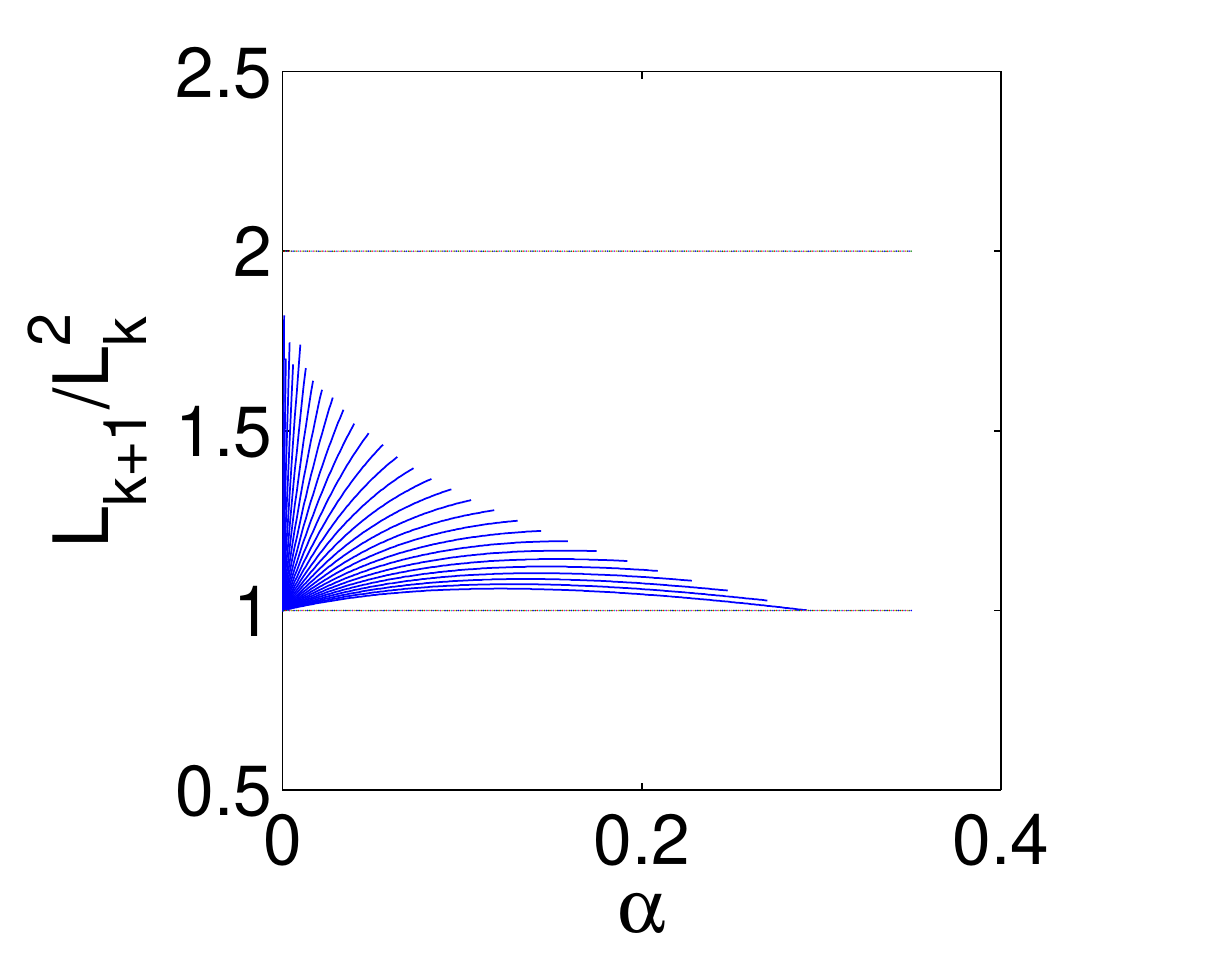}
\caption{Ratio $L_{k+1}/L_k^2$ in $\bigcup _{m=2} ^\infty B_m$. Each line depicts the ratio versus $\alpha$ for a fixed $\beta$.}
\label{fig:ratio1}
\end{figure}

\begin{IEEEproof}
If $(\alpha_k, \beta_k) \in B_m$ for $m\neq 1$, then
\[
\frac{L_{k+1}}{L^2_k} = \frac{1-(1-\alpha_k)^2+\beta_k^2}{(\alpha_k+\beta_k)^2}.
\]

The following calculation establishes the lower bound of the ratio $L_{k+1}/L_{k}^2$:
\begin{align*}
L_{k+1}-L^2_k &=1-(1-\alpha_k)^2+\beta_k^2-(\alpha_k+\beta_k)^2 \\
&=-2\alpha_k^2-2\alpha_k\beta_k+2\alpha_k \\
&= 2\alpha_k(1-(\alpha_k+\beta_k)) \geq 0,
\end{align*}
which holds in $B_m$.

To show the upper bound of the ratio $L_{k+1}/L_{k}^2$, it suffices to prove that
\begin{align*}
L_{k+1}-2L^2_k
&=1-(1-\alpha_k)^2+\beta_k^2-2(\alpha_k+\beta_k)^2 \\
&=-3\alpha_k^2-4\alpha_k\beta_k+2\alpha_k-\beta_k^2 \leq  0.
\end{align*}
The partial derivative with respect to $\beta_k$ is
\[
\frac{\partial{(L_{k+1}-2L^2_k)}}{\partial{\beta_k}}=-2\beta_k-4\alpha_k\leq 0,
\]
which is  non-positive, and so it suffices to consider values on the upper boundary of $B_1$.
\begin{align*}
L_{k+1}-2L^2_k
&=1-(1-\alpha_k)^2+\beta_k^2-2(\alpha_k+\beta_k)^2\\
&=2\beta_k^2-2(\alpha_k+\beta_k)^2\leq 0.
\end{align*}
In consequence, the claimed upper bound on the ratio $L_{k+1}/L_k^2$ holds.
The reader is  referred to Fig.~\ref{fig:ratio1} for a plot of values of $L_{k+1}/L_k^2$.

\end{IEEEproof}

\emph{Proposition 3}:
Suppose that $(\alpha_0, \beta_0) \in B_m$, where $m$ is a positive integer and $m\neq1$. Then,
for $k=1,\ldots, m-1$,
\[
2^k\left(\log L_0^{-1}-\frac{k}{2^k}\right)\leq \log L_k^{-1} \leq 2^k \log L_0^{-1}
\]

\begin{IEEEproof}
From Proposition 2 we have, for $k=0,\ldots, m-2$,
\[
L_{k+1}=a_k L_{k}^2
\]
for some $a_k\in [1,2]$.
Then for $k=1,\ldots, m-1$,
\[
L_k = \left(\prod_{i=1}^{k} a_i\right) L_0^{2^k},
\]
where $a_i \in [1,2]$. Therefore,
\begin{align*}
\log L_k^{-1}= -\left(\sum_{i=1}^{k} \log a_i\right) - \log L_0^{2^{k}}
= -\left(\sum_{i=1}^{k} \log a_i\right) + 2^{k}\log L_0^{-1}.
\end{align*}
Since $\log L_0^{-1}>0$ and
$0 \leq \log a_i \leq 1$ for each $i$, we have
\[
\log L_k^{-1} \leq 2^{k}\log L_0^{-1}.
\]
Finally,
\begin{align*}
\log L_k^{-1}
\geq  -k + 2^{k}\log L_0^{-1}
= 2^{k} \left(\log L_0^{-1} -
\frac{k}{2^{k}}\right).
\end{align*}
\end{IEEEproof}

Suppose that the balanced binary relay tree has $N$ leaf nodes. Then
the height of the fusion center is $\log N$. For convenience, let
$P_N=L_{\log N}$ be (twice) the total error probability at the fusion
center. Substituting $k=\log N$ into Proposition 3, we get the
following result.

\emph{Corollary 1}:
Suppose that $(\alpha_0, \beta_0) \in B_m$, where $m$ is a positive integer and $m\neq1$. If $\log N < m$, then

\[
N\left(\log L_0^{-1}-\frac{\log N}{N}\right)\leq \log P_N^{-1} \leq N \log L_0^{-1}.
\]

Notice that $\log N/N$ tends to zero as $N$ tends to infinity. Hence,
the decay rate of $P_N$ with respect to $N$ is exponential, with error
exponent $\log L_0^{-1}$.  In other words, as $(\alpha_k, \beta_k)$
marches from $B_m$ towards $B_1$, the total error probability at each
level  reduces
exponentially quickly.

\subsection{Case II: Error probability analysis when the system stays inside $R$}
We have derived error probability bounds up until the point where the
trajectory of  the system enters
$B_1$. In this section, we consider the total error probability
redution   from that point on.  First we will establish error
probability bounds for even-height trees. Then we will deduce error
probability bounds for odd-height trees.

\subsubsection{Error probability bounds for even-height trees}

If $(\alpha_0,\beta_0)\in B_m$ for some $m\neq1$, then $(\alpha_{m-1},\beta_{m-1})\in B_1$.
The system afterward stays inside the invariant region $R$ (but not necessarily inside $B_1$).
Hence, the decay rate of the total error probability in the invariant region $R$ determines the asymptotic decay rate.
Without lost of generality, we assume that $(\alpha_0,\beta_0)$ lies in the invariant region $R$. In contrast to Proposition 2, which bounds the ratio $L_{k+1}/L_k^2$, we will bound the ratio $L_{k+2}/L_k^2$ associated with taking two steps.

\emph{Proposition 4}:
Suppose that $(\alpha_k, \beta_k)\in R$. Then,

\[
1\leq \frac{L_{k+2}}{L^2_k} \leq 2.
\]

\begin{figure}[!th]
\begin{center}
\begin{tabular}{cc}
\includegraphics[width=2.6in]{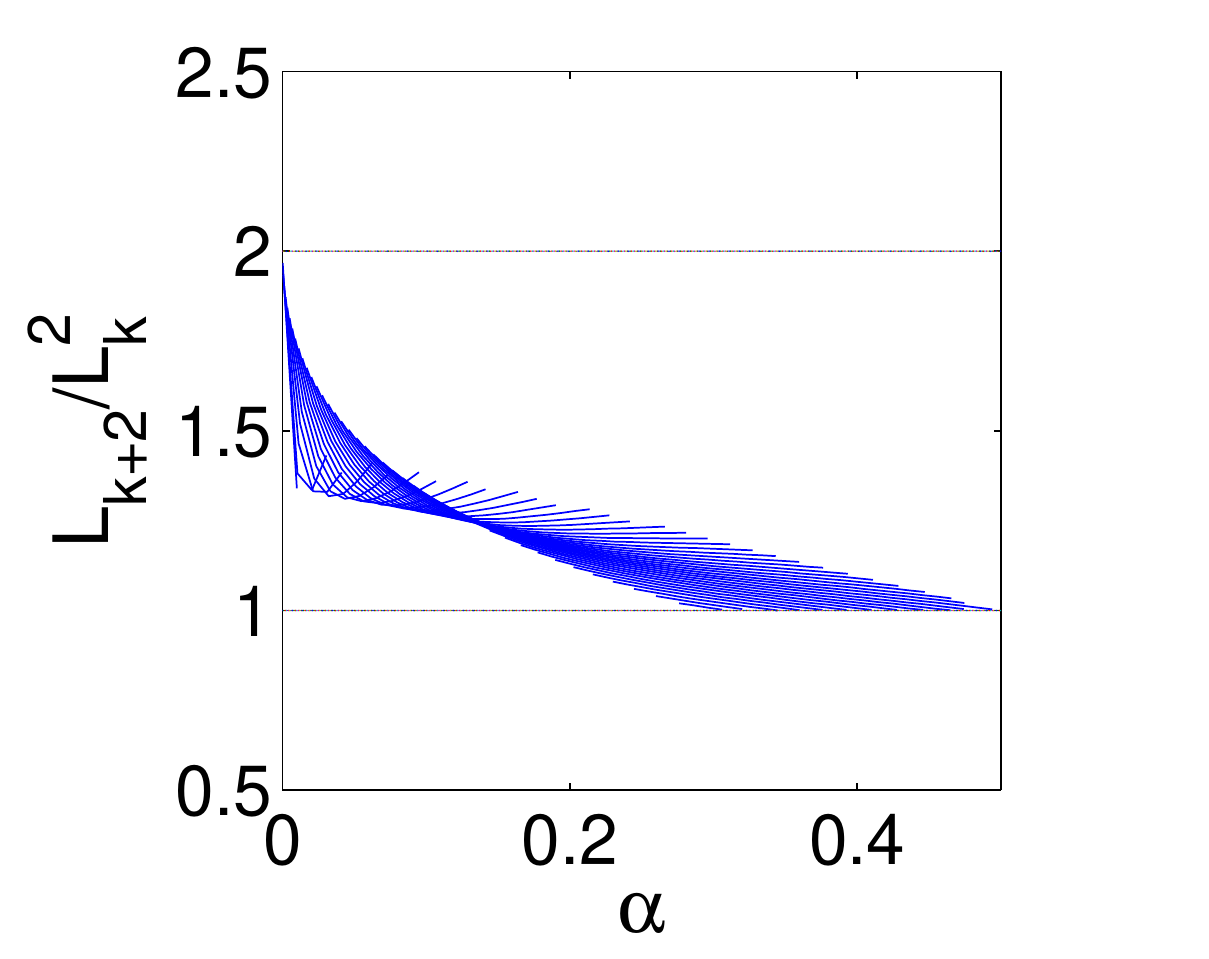} &\includegraphics[width=2.6in]{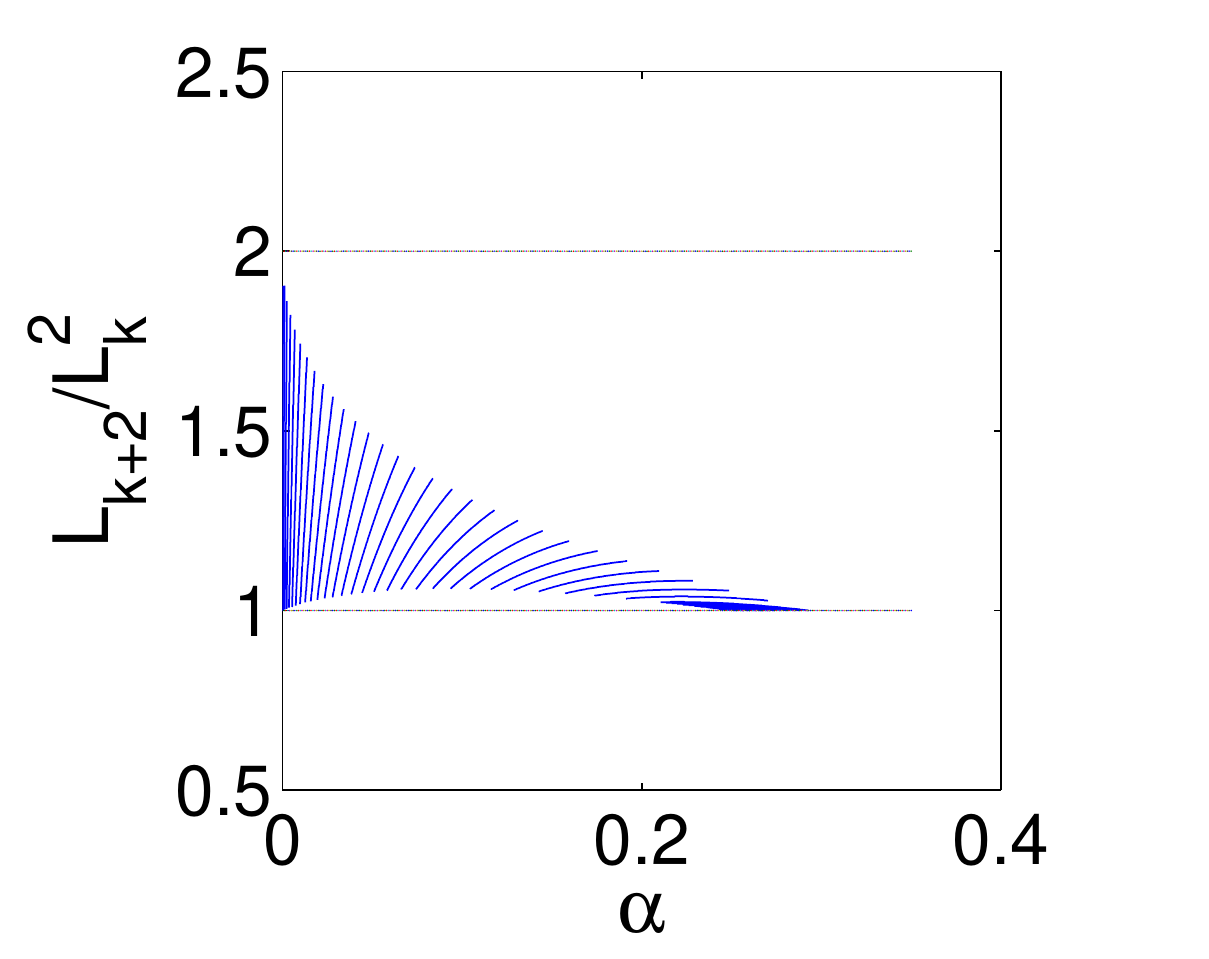}\\
(a)&(b)\\
\end{tabular}
\end{center}
\caption{(a) Ratio $L_{k+2}/L_k^2$ in region $B_1$. (b) Ratio $L_{k+2}/L_{k}^2$ in region $B_2\cap R_\mathcal{U}$. Each line depicts the ratio versus $\alpha$ for a fixed $\beta$.}\label{fig:ratio2}
\end{figure}

\begin{IEEEproof} Because of symmetry, we only have to prove the case
  where $(\alpha_k, \beta_k)$ lies in $R_\mathcal{U}$.  We consider two cases: $(\alpha_k, \beta_k) \in B_1$ and $(\alpha_k, \beta_k) \in B_2\cap R_\mathcal{U}$.

In the first case,
\[
\frac{L_{k+2}}{L^2_k} = \frac{(1-(1-\alpha_k)^2)^2+1-(1-\beta_k^2)^2}{(\alpha_k+\beta_k)^2}.
\]

To prove the lower bound of the ratio, it suffices to show that
\begin{align*}
L_{k+2}-L^2_k
&=(1-(1-\alpha_k)^2)^2+1-(1-\beta_k^2)^2-(\alpha_k+\beta_k)^2 \\ &=(\alpha_k+\beta_k-1)((\alpha_k-\beta_k)^3+2\alpha_k\beta_k(\alpha_k-\beta_k)-(\alpha_k-\beta_k)^2-2\alpha_k^2)\geq0
\end{align*}
which holds for all $(\alpha_k,\beta_k)\in B_1$.

To prove the upper bound of the ratio, it suffices to show that
\[
L_{k+2}-2L^2_k=\alpha_k^4-4\alpha_k^3+2\alpha_k^2-4\alpha_k\beta_k-\beta_k^4\leq0.
\]
The partial derivative with respect to $\beta_k$ is
\[
\frac{\partial{(L_{k+2}-2L^2_k)}}{\partial{\beta_k}}=-4\alpha_k-4\beta_k^3\leq0
\]
which is non-positive. Therefore, it suffices to consider its values
on the curve
$\beta_k=\alpha_k$, on which $L_{k+2}-2L^2_k$ is clearly
non-positive.  The reader can also refer to Fig. \ref{fig:ratio2}(a)
for a plot of values of $L_{k+2}/L_k^2$ in $B_1$.

Now we consider  the second case, namely $(\alpha_k,\beta_k)\in B_2\cap R_\mathcal{U}$, which gives
\[
\frac{L_{k+2}}{L^2 _k} = \frac{1-(1-\alpha_k)^4+\beta_k^4}{(\alpha_k+\beta_k)^2}.
\]

To prove the lower bound of the ratio, it suffices to show that
\begin{align*}
L_{k+2}-L^2_k
&=(1-(1-\alpha_k)^4)+\beta_k^4-(\alpha_k+\beta_k)^2 \\ &=-(\alpha_k+\beta_k-1)(\alpha_k^3-\alpha_k^2\beta_k-3\alpha_k^2+\alpha_k\beta_k^2+2\alpha_k\beta_k-\beta_k^3-\beta_k^2+4\alpha_k)\geq 0.
\end{align*}
Therefore, it suffices to show that
\begin{align*}
\phi(\alpha_k,\beta_k)=&\alpha_k^3-\alpha_k^2\beta_k-3\alpha_k^2+\alpha_k\beta_k^2+2\alpha_k\beta_k-\beta_k^3-\beta_k^2+4\alpha_k\geq 0.
\end{align*}
The partial derivative with respect to $\beta_k$ is
\[
\frac{\partial{\phi}}{\partial{\beta_k}}=-(\alpha_k-\beta_k)^2-2\beta_k^2+2(\alpha_k-\beta_k)\leq 0.
\]
Thus, it is enough to consider the values on the
upper boundaries $\sqrt{1-\beta_k}+\sqrt{\alpha_k} = 1$ and $\alpha_k+\beta_k=1$.

If $\alpha_k+\beta_k=1$, then the inequality is trivial, and if  $\sqrt{1-\beta_k}+\sqrt{\alpha_k} = 1$,
\[
L_{k+2}-L^2 _k=2\alpha_k^2(1-2\sqrt{\alpha_k})(2\alpha_k-6\sqrt{\alpha_k}+5)
\]
and the inequality holds because $\alpha_k \leq \frac{1}{4}$ in region $B_2\cap R_\mathcal{U}$.

The claimed upper bound for the ratio $L_{k+2}/L_k^2$ can be written as
\begin{align*}
L_{k+2}-2L^2_k
&=(1-(1-\alpha_k)^4)+\beta_k^4-2(\alpha_k+\beta_k)^2\\
&=-\alpha_k^4+4\alpha_k^3-8\alpha_k^2+4\alpha_k-4\alpha_k\beta_k+\beta_k^4-2\beta_k^2 \leq 0.
\end{align*}
The partial derivative with respect to $\beta_k$ is
\[
\frac{\partial{(L_{k+2}-2L^2_k)}}{\partial{\beta_k}}=-4\alpha_k+4\beta_k^3-4\beta_k\leq0.
\]
Again, it is sufficient  to consider values on the upper boundary of $B_1$. Therefore,
\begin{align*}
L_{k+2}-2L^2_k=2\beta_k^2-2(\alpha_k+\beta_k)^2 \leq 0.
\end{align*}
The reader is  referred to Fig. \ref{fig:ratio2}(b) for a plot of values of $L_{k+2}/L_k^2$ in $B_2\cap R_\mathcal{U}$.

\end{IEEEproof}

Proposition 4 gives  bounds on  the  relationship between $L_k$ and
$L_{k+2}$ in the invariant region $R$. Hence, in the special case of
trees with even height,  that is, when $\log N$ is an even integer, it is easy
to bound $P_N$ in terms of $L_0$. In fact, we will bound $\log
P_N^{-1}$ which in turn provides bounds for $P_N$.

\begin{Theorem}
If $(\alpha_0,\beta_0)$ is in the invariant region $R$ and $\log N$ is even, then
\[
\sqrt{N} \left(\log L_0^{-1} - \frac{\log{\sqrt{N}}}{\sqrt{N}}\right)
\leq \log P_N^{-1} \leq \sqrt{N}\log L_0^{-1}.
\]
\end{Theorem}

\begin{IEEEproof} If $(\alpha_0, \beta_0) \in R$, then we have
  $(\alpha_k, \beta_k)\in R$ for $k=0,1,\ldots,\log N-2$. From
  Proposition 4, we have
\[
L_{k+2}=a_k  L_{k}^2
\]
for $k=0,1,\ldots,\log N-2$ and some $a_k \in [1,2]$. Therefore, for
$k=2,4,\ldots, \log N $, we have
\[
L_k = \left(\prod_{i=1}^{k/2} a_i\right) L_0^{2^{k/2}},
\]
where $a_i \in [1,2]$. Substituting $k=\log N$, we have
\[
P_N
= \left(\prod_{i=1}^{\lsN} a_i\right) L_0^{2^{\lsN}}
= \left(\prod_{i=1}^{\lsN} a_i\right) L_0^{\sqrt{N}}.
\]
Hence,
\begin{align*}
\log P_N^{-1}
&= -\left(\sum_{i=1}^{\lsN} \log a_i\right) + \sqrt{N}\log L_0^{-1}.
\end{align*}
Notice that $\log L_0^{-1}>0$ and,  for each $i$,
$0 \leq \log a_i \leq 1$. Thus,
\[
\log P_N^{-1} \leq \sqrt{N}\log L_0^{-1}.
\]
Finally,
\begin{align*}
\log P_N^{-1}
\geq  -\lsN+ \sqrt{N}\log L_0^{-1}
= \sqrt{N} \left(\log L_0^{-1} -
\frac{\lsN}{\sqrt{N}}\right).
\end{align*}
\end{IEEEproof}

\subsubsection{Error probability bounds for odd-height trees}

Next we explore the case of trees with odd height, i.e., $\log N$ is
an odd integer. Assume that $(\alpha_0, \beta_0)$ lies in the
invariant region $R$. First, we will establish general bounds for
odd-height trees. Then we deduce bounds for the case where there
exists $(\alpha_k, \beta_k)\in B_2\cap R_\mathcal{U}$ for some $k\in
\{0,1,\ldots,\log N-1\}$.

For odd-height trees, we need to know how much the total error
probability is reduced by moving up one level in the tree.

\emph{Proposition 5}:
If $(\alpha_k, \beta_k) \in \mathcal{U}$, then
\[
1\leq \frac{L_{k+1}}{L^2_k}
\]
and
\[
\frac{L_{k+1}}{L_k} \leq 1.
\]

\begin{figure}[!th]
\begin{center}
\begin{tabular}{cc}
\includegraphics[width=2.6in]{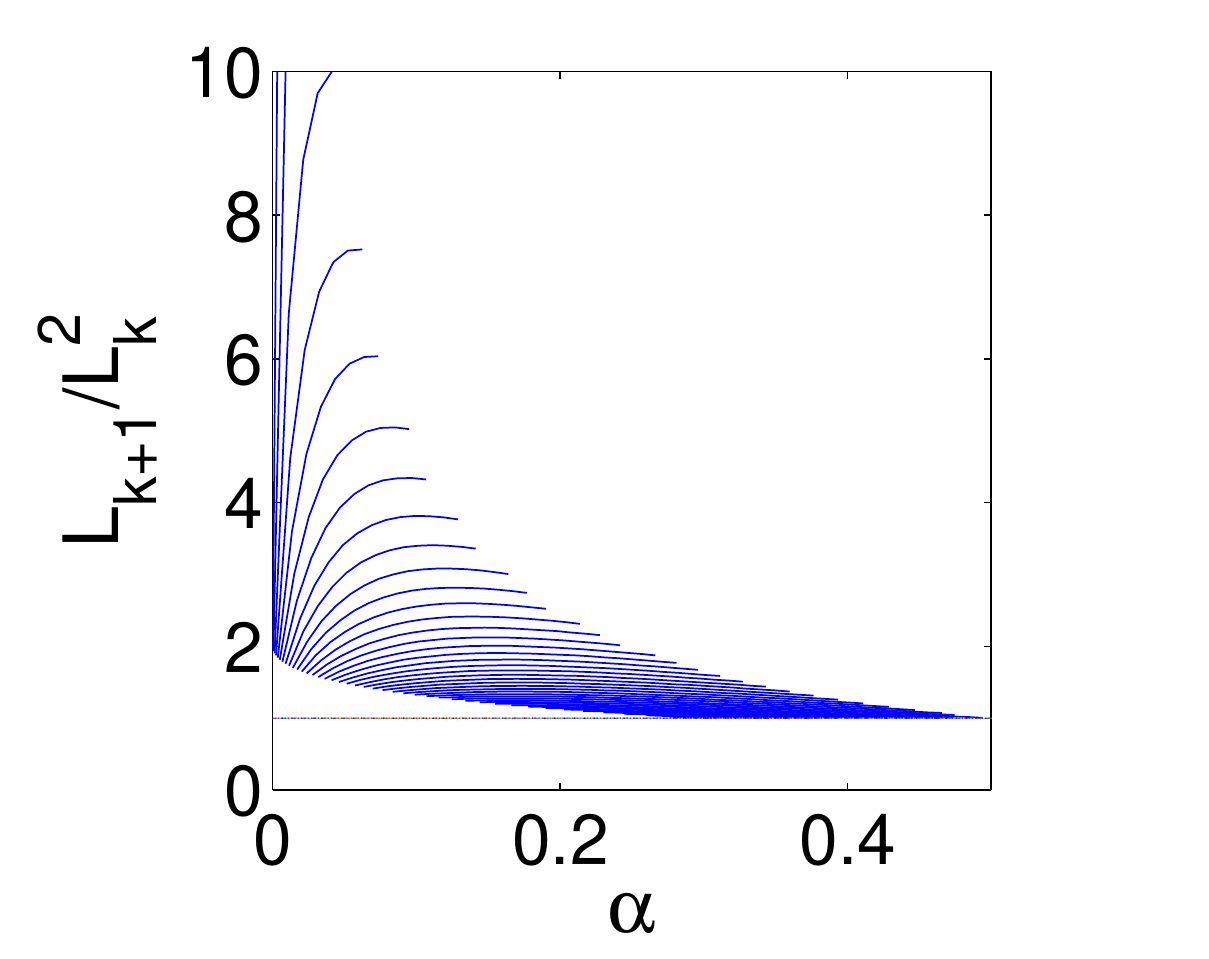}& \includegraphics[width=2.6in]{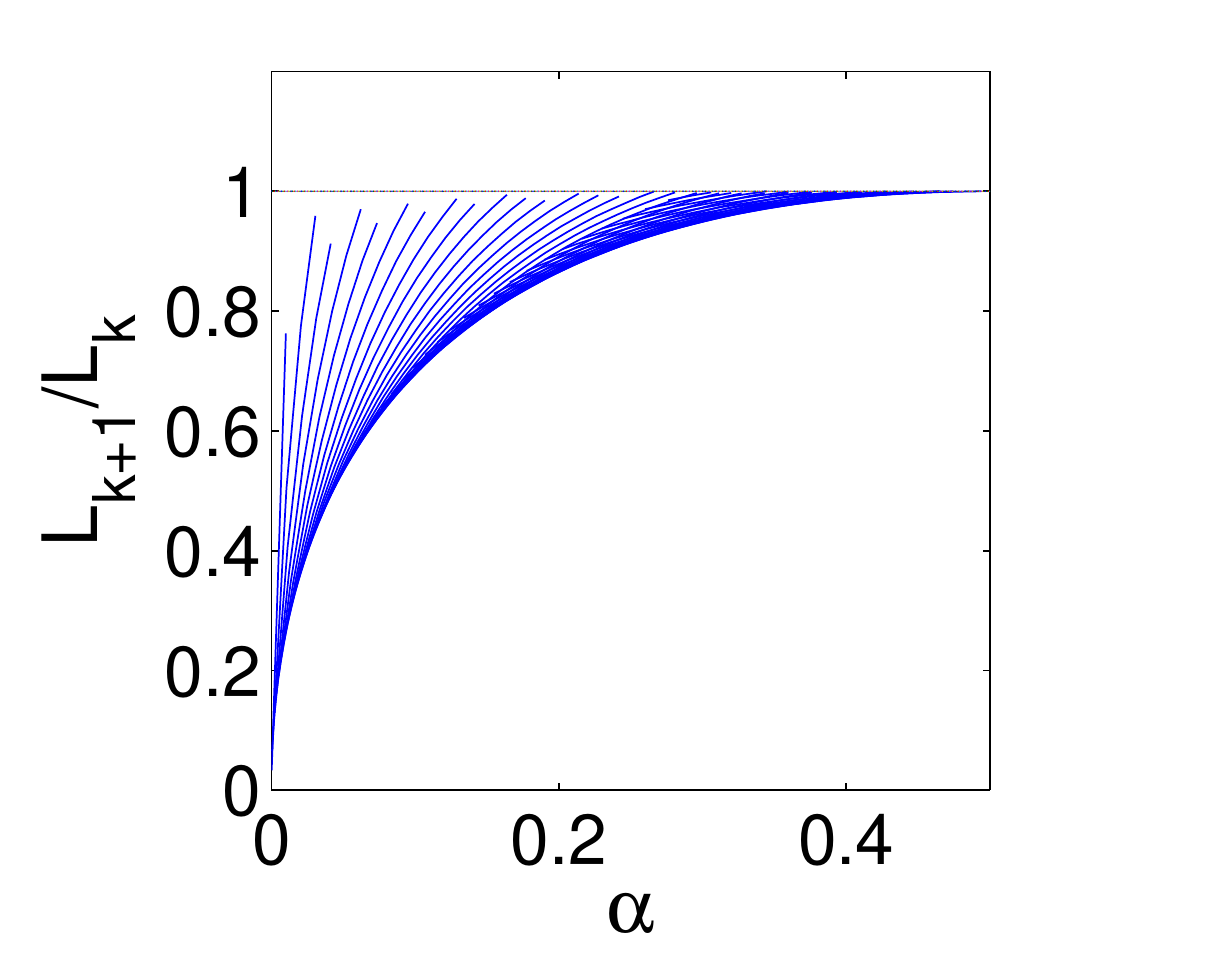}\\
(a)&(b)\\
\end{tabular}
\end{center}
\caption{(a) Ratio $L_{k+1}/L_k^2$ in region $\mathcal{U}$. (b) Ratio $L_{k+1}/L_k$ in region $\mathcal{U}$. Each line depicts the ratio versus $\alpha$ for a fixed $\beta$.}\label{fig:ratio5}
\end{figure}

\begin{IEEEproof}
The first inequality is equivalent to
\begin{align*}
L_{k+1}-L^2 _k
&=1-(1-\alpha_k)^2+\beta_k^2-(\alpha_k+\beta_k)^2\\
&= 2\alpha_k(1-(\alpha_k+\beta_k)) \geq  0,
\end{align*}
which holds for all $(\alpha_k, \beta_k) \in \mathcal{U}$.

The second inequality  is equivalent to
\begin{align*}
L_{k+1}-L_k
&=1-(1-\alpha_k)^2+\beta_k^2-(\alpha_k+\beta_k)\\
&=(\alpha_k-\beta_k)(1-(\alpha_k+\beta_k)) \leq  0,
\end{align*}
which holds for all $(\alpha_k, \beta_k) \in \mathcal{U}$.
Fig. \ref{fig:ratio5} gives a  plot of values of $L_{k+1}/L_k^2$ and $L_{k+1}/L_k$ in $\mathcal{U}$.

\end{IEEEproof}

Using  Propositions 4 and 5, we are about to calculate error
probability bounds for odd-height trees as follows.
\begin{Theorem}
If $(\alpha_0, \beta_0)\in R$ and $\log N$ is odd, then
\begin{align*}
\sqrt{\frac{N}{2}} \left(\log L_0^{-1} - \frac{\log{\sqrt{\frac{N}{2}}}}{\sqrt{\frac{N}{2}}}\right)
\leq \log P_N^{-1} \leq \sqrt{2N}\log L_0^{-1}.
\end{align*}
\end{Theorem}

\begin{IEEEproof}
By Proposition 5, we have
\[
L_{1}=\widetilde{a} L_{0}^2
\]
for some $\widetilde{a} \geq 1$.
And, by Proposition 4, the following identity holds.
\[
L_{k+2}=a_k L_{k}^2
\]
for $k=1,3,\ldots,\log N-2$ and some $a_k \in [1,2]$. Hence, we can write
\[
L_{k}=\widetilde{a}\left(\prod_{i=1}^{(k-1)/2} a_i\right) L_0^{2^{(k+1)/2}},
\]
where $1\leq a_i\leq2$ for $i=1,2,\ldots,(k-1)/2$ and $\widetilde{a}\geq 1$.
Let $k=\log N$, we have
\[
P_N
= \widetilde{a}\left(\prod_{i=1}^{\log{\sqrt{\frac{N}{2}}}} a_i\right) L_0^{2^{\log{\sqrt{2N}}}}
= \widetilde{a}\left(\prod_{i=1}^{\log{\sqrt{\frac{N}{2}}}} a_i\right) L_0^{\sqrt{2N}},
\]
and so
\begin{align*}
\log P_N^{-1}
&= -\log \widetilde{a}-\left(\sum_{i=1}^{\log{\sqrt{\frac{N}{2}}}} \log a_i\right) + \sqrt{2N}\log L_0^{-1}.
\end{align*}
Notice that $\log L_0^{-1}>0$ and for each $i$,
$ \log a_i \geq 0$. Moreover, $ \log\widetilde{a}\geq0 $. Hence,
\[
\log P_N^{-1} \leq \sqrt{2N}\log L_0^{-1}.
\]

It follows by  Proposition 5 that
\[
L_{1}=\widetilde{a} L_{0}
\]
for  some $\widetilde{a} \leq 1$. Thus,
\[
L_{k}=\widetilde{a}\left(\prod_{i=1}^{(k-1)/2} a_i\right) L_0^{2^{(k-1)/2}},
\]
where $1\leq a_i\leq2$ for $i=1,2\ldots,(k-1)/2$ and $ \widetilde{a}\leq1$.
Hence,
\[
P_N
= \widetilde{a}\left(\prod_{i=1}^{\log{\sqrt{\frac{N}{2}}}} a_i\right) L_0^{2^{\log{\sqrt{\frac{N}{2}}}}}
= \widetilde{a}\left(\prod_{i=1}^{\log{\sqrt{\frac{N}{2}}}} a_i\right) L_0^{\sqrt{\frac{N}{2}}}
\]
and so
\begin{align*}
\log P_N^{-1}
&= -\log \widetilde{a}-\left(\sum_{i=1}^{\log{\sqrt{\frac{N}{2}}}} \log a_i\right) + \sqrt{\frac{N}{2}}\log L_0^{-1}.
\end{align*}

Notice that $\log L_0^{-1}>0$ and for each $i$, $0 \leq \log a_i \leq 1 $ and $\log {\widetilde{a}} \leq 0 $. Thus,
\begin{align*}
\log P_N^{-1}
\geq -\log{\sqrt{\frac{N}{2}}} + \sqrt{\frac{N}{2}}\log L_0^{-1}
= \sqrt{\frac{N}{2}} \left(\log L_0^{-1} -
\frac{\log{\sqrt{\frac{N}{2}}}}{\sqrt{\frac{N}{2}}}\right).
\end{align*}

\end{IEEEproof}

Now we consider the special case where there exists
$k\in\{0,1,\ldots,\log N -1\}$ such that $(\alpha_k,\beta_k)\in
B_2\cap R_\mathcal{U}$.

\begin{figure}[htbp]
\centering
\includegraphics[width=3in]{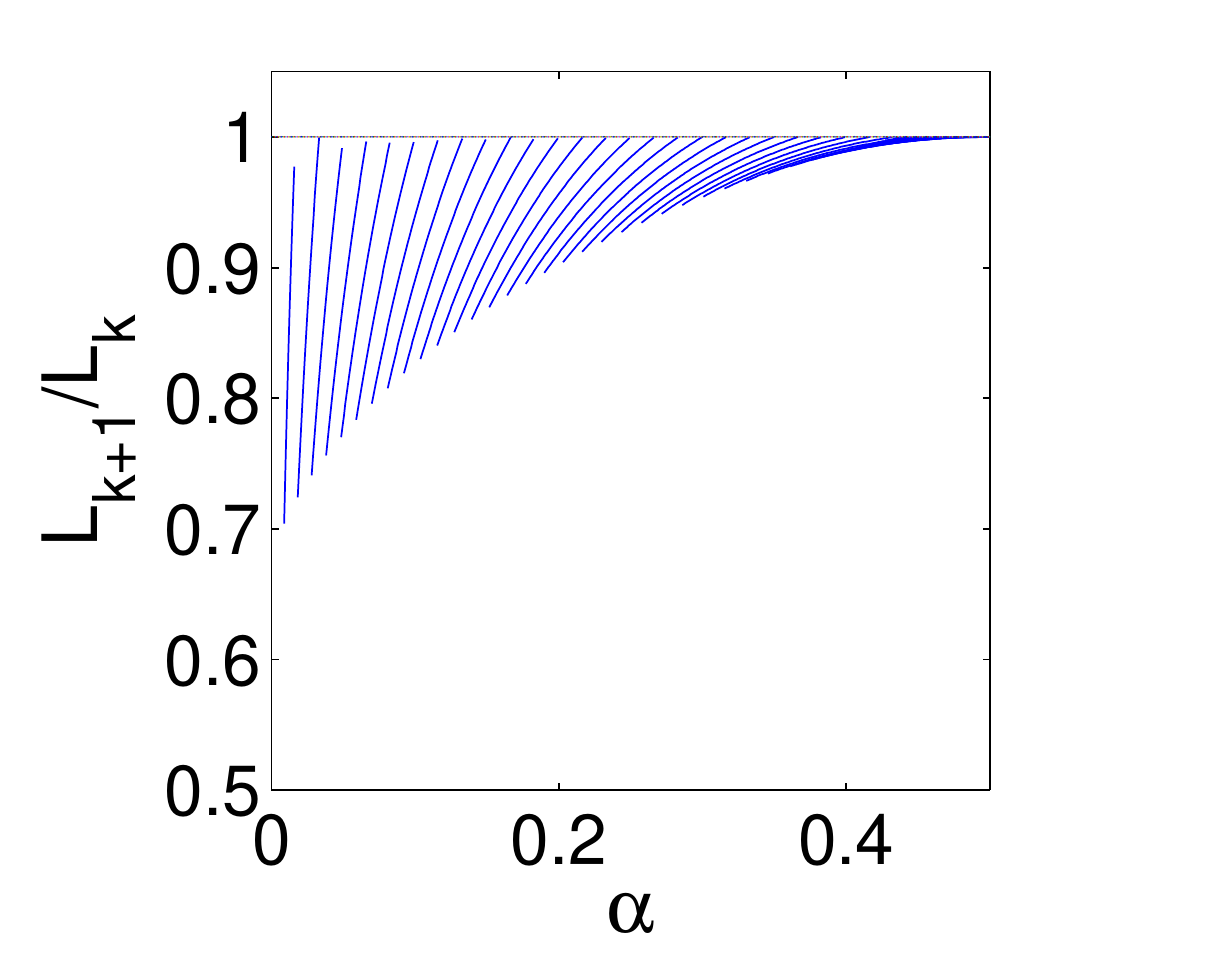}
\caption{Ratio $L_{k+1}/L_k$ in the region $f(B_2\cup R_\mathcal{U})$. Each line depicts the ratio versus $\alpha$ for a fixed $\beta$.}
\label{fig:ratio4}
\end{figure}

\emph{Proposition 6}:
Suppose that $(\alpha_k, \beta_k) \in B_1$ and $(\alpha_{k-1}, \beta_{k-1}) \in B_2\cap R_\mathcal{U}$. Then,
\[
\frac{1}{2}\leq \frac{L_{k+1}}{L_k} \leq 1.
\]
\begin{IEEEproof}
The right inequality is trivial.
By Proposition 2, if $(\alpha_{k-1}, \beta_{k-1})\in B_2 \cap R_\mathcal{U}$, then
\[
1\leq \frac{L_{k}}{L^2_{k-1}} \leq 2,
\]
i.e.,
\[
\frac{1}{2}\leq \frac{L^2_{k-1}}{L_{k}} \leq 1,
\]
and in consequence of Proposition 4, if $(\alpha_{k-1}, \beta_{k-1})\in B_2 \cap R_\mathcal{U}$, then
\[
1\leq \frac{L_{k+1}}{L^2_{k-1}} \leq 2.
\]
Therefore,
\[
\frac{1}{2}\leq \frac{L_{k+1}}{L_k}.
\]
In this case,  Fig. \ref{fig:ratio4} gives a  plot of the ratio $L_{k+1}/L_k$ in the region $f(B_2\cup R_\mathcal{U})$.

\end{IEEEproof}

We have proved in Proposition 4 that if $(\alpha_k, \beta_k)$ is in
$B_2\cap R_\mathcal{U}$, then the ratio $L_{k+2}/L^2_k \in
[1,2]$. However if we analyze each level of fusion, it can be seen
that the total error probability decreases exponentially fast from
$B_2\cap R_\mathcal{U}$ to $B_1$ (Proposition 2). Proposition 6 tells
us that the fusion from $B_1$ to $R_\mathcal{L}$ is a bad step, which
does not contribute significantly  in decreasing the total error probability.

We can now provide bounds for the total error probability at the fusion center.

\begin{Theorem}
  Suppose that $(\alpha_0, \beta_0)\in R$, $\log N$ is an odd integer,
  and there exists $k\in \{0, 1, \ldots, \log N-1\}$ such that
  $(\alpha_k, \beta_k) \in B_2\cap R_\mathcal{U}$.

If $k$ is even, then
\begin{align*}
\sqrt{2N} \left(\log L_0^{-1} -  \frac{\log{\sqrt{2N}}}{\sqrt{2N}}\right)
\leq \log P_N^{-1} \leq \sqrt{2N}\log L_0^{-1}.
\end{align*}

If $k$ is odd, then
\begin{align*}
\sqrt{\frac{N}{2}} \left(\log L_0^{-1} -  \frac{\log{\sqrt{\frac{N}{2}}}}{\sqrt{\frac{N}{2}}}\right)
\leq \log P_N^{-1} \leq \sqrt{\frac{N}{2}}\log L_0^{-1}+1.
\end{align*}
\end{Theorem}
\begin{IEEEproof}
If $(\alpha_m, \beta_m)\in B_2\cap R_\mathcal{U}$ and $m$ is even, then by Proposition 2, we have

\[
L_{m+1}=\widetilde{a} L_m^2
\]
for some $\widetilde{a} \in [1,2]$.

By Proposition 4, we have
\[
L_{k+2}=a_k L_{k}^2
\]
for $k=0,2,\ldots,m-2,m+1,\ldots,\log N-2$, and some $a_k\in [1,2]$.
Hence,
\[
L_{k}=\left(\prod_{i=1}^{(k+1)/2} a_i\right) L_0^{2^{(k+1)/2}},
\]
where $a_i \in [1,2]$.

Let $k=\log N$, we have
\[
P_N=L_{\log N}
= \left(\prod_{i=1}^{\log \sqrt {2N}} a_i\right) L_0^{\sqrt{2N}}.
\]
Therefore,
\begin{align*}
\log P_N^{-1}
&= -\left(\sum_{i=1}^{\log{\sqrt{2N}}} \log a_i\right) + \sqrt{2N}\log L_0^{-1}.
\end{align*}

Notice that $\log L_0^{-1}>0$ and for each $i$,
$0 \leq \log a_i \leq 1$. Thus,
\[
\log P_N^{-1} \leq \sqrt{2N}\log L_0^{-1}.
\]
Finally,
\begin{align*}
\log P_N^{-1}
\geq  -\log{\sqrt{2N}} + \sqrt{2N}\log L_0^{-1}
= \sqrt{2N} \left(\log L_0^{-1} -
\frac{\log{\sqrt{2N}}}{\sqrt{2N}}\right).
\end{align*}

If $(\alpha_m, \beta_m)\in B_2\cap R_\mathcal{U}$ and $m$ is odd, then by Proposition 6 we have
\[
L_{m+2}=\widetilde{a} L_{m+1}
\]
for some $\widetilde{a} \in [1/2, 1]$.

It follows from Proposition 4 that
\[
L_{k+2}=a_k L_{k}^2
\]
for $k=0,2,\ldots,m-1,m+2,\ldots,\log N-2$ and some $a_k\in [1,2]$.
Therefore,
\[
L_{k}=\widetilde{a}\left(\prod_{i=1}^{(k-1)/2} a_i\right)  L_0^{2^{(k-1)/2}},
\]
where $1\leq a_i\leq2$ for $i=1,2,\ldots,(k-1)/2$ and $1/2\leq \widetilde{a}\leq1$.
Hence,
\[
P_N
= \widetilde{a}\left(\prod_{i=1}^{(\log N-1)/2} a_i\right) L_0^{2^{(\log N-1)/2}}
= \widetilde{a}\left(\prod_{i=1}^{\log \sqrt{\frac{N}{2}}} a_i\right) L_0^{\sqrt{\frac{N}{2}}}
\]
and so
\begin{align*}
\log P_N^{-1}
&= -\log \widetilde{a}-\left(\sum_{i=1}^{\log \sqrt{\frac{N}{2}}} \log a_i\right) + \sqrt{\frac{N}{2}}\log L_0^{-1}.
\end{align*}

Notice that $\log L_0^{-1}>0$ and for each $i$,
$0 \leq \log a_i \leq 1$ and $-1\leq \log{\widetilde{a}}\leq0$. Thus,
\[
\log P_N^{-1} \leq \sqrt{\frac{N}{2}}\log L_0^{-1}+1.
\]
Finally,
\begin{align*}
\log P_N^{-1}
\geq -\log{\sqrt{\frac{N}{2}}} + \sqrt{\frac{N}{2}}\log L_0^{-1}
= \sqrt{\frac{N}{2}} \left(\log L_0^{-1} -
\frac{\log{\sqrt{\frac{N}{2}}}}{\sqrt{\frac{N}{2}}}\right).
\end{align*}

\end{IEEEproof}

Finally, by combining all of the analyses above for step-wise
reduction  of the total error probability, we can write general bounds when the initial error probability pair $(\alpha_0,\beta_0)$ lies inside $B_m$, where $m\neq 1$.
\begin{Theorem}
Suppose that $(\alpha_0, \beta_0) \in B_m$, where $m$ is an integer and $m\neq1$.

If $\log N \leq m-1$, then\space(Theorem 1)
\[
N \left(\log L_0^{-1} - \frac{\log{N}}{N}\right)
\leq \log P_N^{-1} \leq N\log L_0^{-1}.
\]

If $\log N>m-1$, and $\log N-m$ is odd, then
\begin{align*}
\sqrt{2^{m-1}N} \left(\log L_0^{-1} -  \frac{\log{\sqrt{2^{m-1}N}}}{\sqrt{2^{m-1}N}}\right)
\leq \log P_N^{-1} \leq \sqrt{2^{m-1}N}\log L_0^{-1}.
\end{align*}

If $\log N>m-1$, and $\log N-m$ is even, then
\begin{align*}
\sqrt{2^{m-2}N} \left(\log L_0^{-1} - \frac{\log{\sqrt{2^{m-2}N}}}{\sqrt{2^{m-2}N}}\right)\leq \log P_N^{-1} \leq \sqrt{2^{m}N}\log L_0^{-1}.
\end{align*}

\end{Theorem}
The proof uses similar arguments as that of Theorem 2 and it is provided in Appendix A.

\subsection{Invariant region in $B_1$}
\begin{figure}[htbp]
\centering
\includegraphics[width=3.5in]{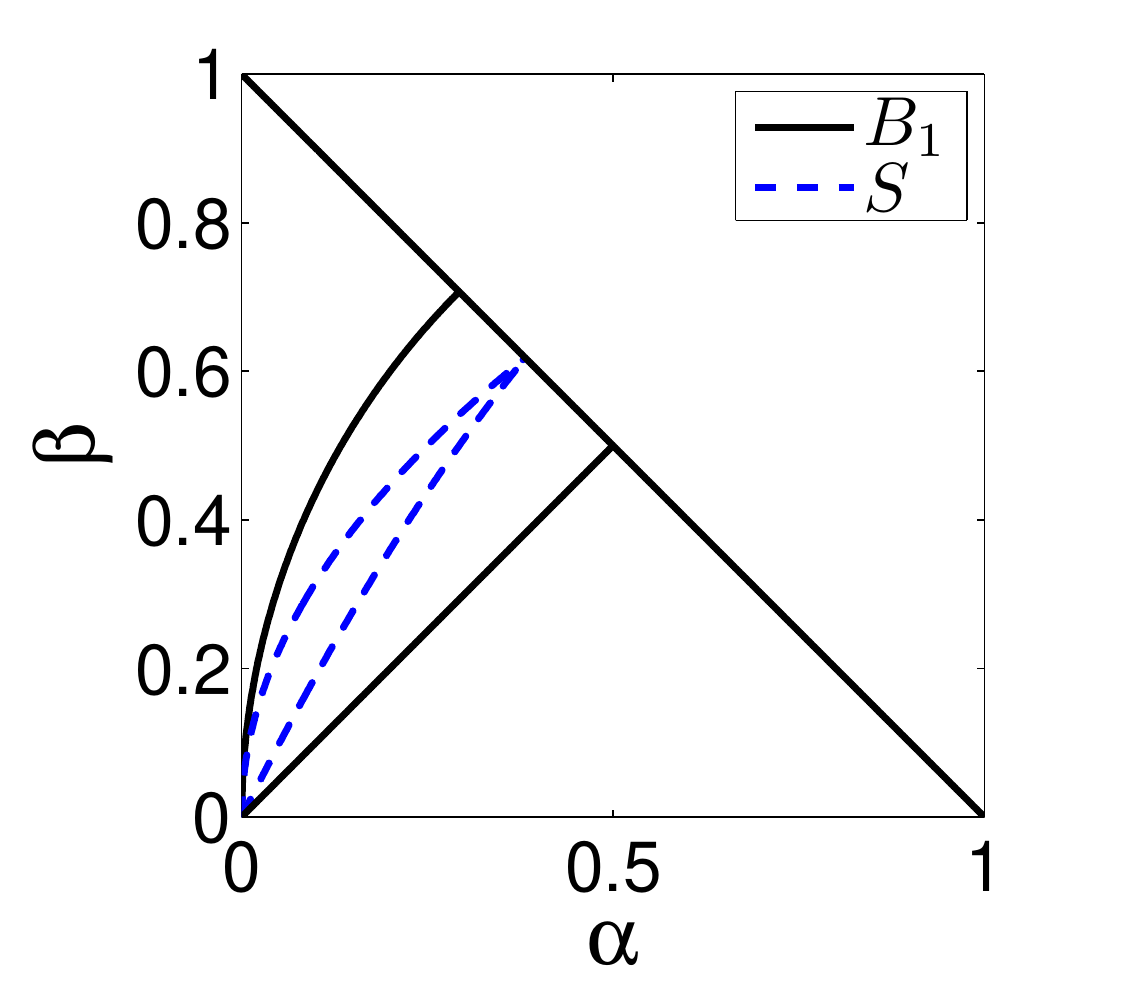}
\caption{Invariant region $S$ (between dashed lines) lies inside $B_1$ (between solid lines).}
\label{fig:s}
\end{figure}

Consider the region $\{(\alpha,\beta)\in
\mathcal{U}|\beta\leq\sqrt{\alpha} \text{ and }
\beta\geq1-(1-\alpha)^2\}$, which  is a subset of $B_1$ (see
Fig. \ref{fig:s}). Denote the union of this  region and its reflection with respect to $\beta=\alpha$ by $S$.
It turns out that $S$ is also invariant.

\emph{Proposition 7}:
If $(\alpha_{k_0},\beta_{k_0})\in S$, then $(\alpha_k,\beta_k) \in S$ for all $k\geq k_0$.

The proof is given in Appendix B.

We have given bounds for $P_N$, which is (twice) the total error
probability. It turns out that for the case where
$(\alpha_0,\beta_0)\in S$, we can bound Type I and Type II errors
individually.

\emph{Proposition 8}: If $(\alpha_k,\beta_k)\in S$, then
\[
1\leq\frac{\alpha_{k+2}}{\alpha_k^2}\leq 4
\]
and
\[
1\leq\frac{\beta_{k+2}}{\beta_k^2}\leq 4.
\]

\emph{Remark}: It is easy to see that as long as the system stays
inside $B_1$, then in a similar vein, these ratios $\alpha_{k+2}/\alpha_k^2$
and $\beta_{k+2}/\beta_k^2$ are lower bounded by 1 and upper bounded
by a constant. But recall that $B_1$ is not an invariant region. Thus,
it is more interesting to consider $S$.

Proofs are omitted because they along similar lines to those in the
other proofs. As before, these inequalities give rise to bounds on
sequences $\{\alpha_k\}$ and $\{\beta_k\}$. For example, for
$\{\alpha_k\}$, we have the following.

\emph{Corollary 2}:
If $(\alpha_0, \beta_0)\in S$, then for even $k$ we have
\[
2^{k/2}\left(\log \alpha_0^{-1}-\frac{k}{2^{k/2}}\right)\leq \log \alpha_k^{-1} \leq 2^{k/2} \log \alpha_0^{-1}.
\]

We have derived error probability bounds for balanced binary relay trees under several scenarios.
In the next section, we will use these bounds to study the asymptotic rate of convergence.

\section{Asymptotic Rates}

The  asymptotic decay rate of the
total error probability with respect to $N$ is considered  while
the performance of the sensors is constant is the first problem to be tackled.  Then we allow the sensors
to be asymptotically crummy, in the sense that
$\alpha_0+\beta_0\rightarrow 1$. We prove that the total error
probability still converges to $0$ under certain conditions.

\subsection{Asymptotic decay rate}

Notice that as  $N$ becomes large, the sequence $\{(\alpha_k,
\beta_k)\}$ will eventually move into the invariant region $R$ at some level and stays
inside from that point. Therefore, it suffices to consider the decay rate in
the invariant region $R$. Because error probability bounds for trees
with odd height differ from those of the even-height tree by a
constant term, without loss of generality, we will only consider trees with even height.

\emph{Proposition 9}: If $L_0=\alpha_0+\beta_0$ is fixed, then
\begin{equation*}
\log P_N^{-1} \sim \log L_0^{-1}\sqrt{N}.
\end{equation*}

\begin{IEEEproof} If $L_0=\alpha_0+\beta_0$ is fixed, then by Theorem 1 we immediately see that $P_N\to 0$ as $N\to\infty$ ($\log P_N^{-1}\to\infty$) and
\[
1-\frac{\log{\sqrt N}}{\log L_0^{-1}\sqrt N} \leq\frac{\log P_N^{-1}}{\log L_0^{-1}\sqrt{N}} \leq 1.
\]
In addition, because $\log{\sqrt{N}}/\sqrt{N}\to 0$, we have
\[
\frac{\log P_N^{-1}}{\log L_0^{-1}\sqrt{N}}\rightarrow 1,
\]
which means
\begin{equation*}
\log P_N^{-1} \sim \log L_0^{-1}\sqrt{N}.
\end{equation*}
\end{IEEEproof}
This implies that the convergence of the total error probability is
sub-exponential, more precisely, the decay exponent is essentially
$\sqrt N$.

Given $L_0\in(0,1)$ and $\ep\in (0,1)$, suppose that we wish to determine
how many sensors we need to have so that $P_N\leq\ep$. The solution is
simply to find an
$N$ (e.g., the smallest) satisfying the inequality
\[
\sqrt{N} \left(\log L_0^{-1} - \frac{\log{\sqrt{N}}}{\sqrt{N}}\right)
\geq -\log\ep.
\]
The smallest $N$ grows like $\Theta((\log\ep)^2)$ (cf., \cite{Gubner},
in which the smallest $N$ has a larger growth rate).

\subsection{Crummy sensors}
In this part we allow the total error probability of each sensor,
denoted by $L_0^{(N)}$, to depend on $N$ but still to be constant
across sensors.

If $L_0^{(N)}$ is bounded by some constant $L \in (0,1)$ for all $N$, then clearly $P_N\to 0$.
It is more interesting to consider $L_0^{(N)}\to 1$, which means that sensors are asymptotically crummy.

\emph{Proposition 10}:
Suppose that $L_0^{(N)}\to 1$, or specifically write $L_0^{(N)}=1-\eta_N$ with
$\eta_N\to 0$.
Then, $P_N\rightarrow 0$ if and only if $\eta_N=\omega(1/\sqrt{N})$.

\begin{IEEEproof}
For
sufficiently large $N$,
\[
\sqrt{N}  \frac{\log (L_0^{(N)})^{-1}}{2}
\leq \log P_N^{-1} \leq \sqrt{N}\log (L_0^{(N)})^{-1}.
\]
We conclude that $P_N\to 0$ if and only if
\[
\sqrt{N}\log (L_0^{(N)})^{-1}\to \infty.
\]
Therefore,
\[
\sqrt{N}\log (L_0^{(N)})^{-1} = -\sqrt{N}\log (1-\eta_N).
\]
But as $x\to 0$, $-\log(1-x)\sim x/\ln(2)$. Hence,
$P_N\to 0$ if and only if $\eta_N\sqrt{N}\to \infty$,
often written $\eta_N=\omega(1/\sqrt{N})$.

\end{IEEEproof}
Now suppose that $c_1/\sqrt{N}\leq \eta_N\leq c_2/\sqrt{N}$. In
this case, for large $N$ we deduce that
\[
c_1\leq \log P_N^{-1}\leq c_2,
\]
or equivalently,
\[
2^{-c_2}\leq P_N\leq 2^{-c_1}.
\]

Finally, if $\eta_N=o(1/\sqrt{N})$ (i.e., $\eta_N$ converges to $0$
strictly faster than $1/\sqrt{N}$), then $P_N\to 1$.

\section{Conclusion}
We have studied the detection performance of a balanced binary relay
tree of sensors and fusion nodes. We precisely describe the evolution of error probabilities in
the $(\alpha, \beta)$ plane as we move up the tree. This allows us to deduce error probability bounds
at the fusion center as functions of $N$ under several different
scenarios.  These bounds imply that the total error probability
converges to $0$ sub-linearly, with a decay exponent that is
essentially $\sqrt N$. In addition, we allow all sensors to be
asymptotically crummy, in which case we deduce the necessary and
sufficient condition for the total error probability to converge to
$0$. All our results apply not only  to the fusion center, but also to
any other node in the tree network. In other words, we can similarly
analyze a sub-tree inside the original tree network.

Needless to say, our conclusions are subject to our particular
architecture and assumptions. A series of questions follows:
Considering the same tree configuration, how does the error
probability behave if prior probabilities are not equal?  Considering
balanced binary relay trees with sensor and/or connection failures,
how would the error probability behave? More generally, what can we
say about unbalanced relay trees?  We also would like to apply our
methodology to other configurations, such as a tandem structure.

\appendices
\section{Proof of Theorem 4}
If $\log N \leq m-1$, then this scenario is the same as that of  Theorem 1.
Therefore,
\[
N \left(\log L_0^{-1} - \frac{\log{N}}{N}\right)
\leq \log P_N^{-1} \leq N\log L_0^{-1}.
\]

If $\log N>m-1$ and $\log N-m$ is odd, then it takes $(m-1)$ steps for
the system to move into $B_1$. After it arrives in $B_1$, there is an
even number of
levels left because $\log N-m$ is odd.

By Proposition 2, we have
\[
L_{k+1}=\widetilde{a_k} L_k^2
\]
for $k=0,1,\ldots,m-2$ and some $\widetilde{a_k} \in [1,2]$,
and in consequence of Proposition 4,
\[
L_{k+2}=a_k L_k^2
\]
for $k=m-1,m-3,\ldots,\log N-2$ and some $a_k \in [1,2]$.
Thus,
\[
L_{k}=\left(\prod_{i=1}^{m-1+\frac{k-(m-1)}{2}} a_i\right) L_0^{2^{m-1+\frac{k-(m-1)}{2}}},
\]
where $a_i \in [1,2]$.

Let $k=\log N$. Then we obtain
\begin{align*}
P_N= \left(\prod_{i=1}^{\log{\sqrt{2^{m-1}N}}} a_i\right) L_0^{2^{\log{\sqrt{2^{m-1}N}}}}
= \left(\prod_{i=1}^{\log{\sqrt{2^{m-1}N}}} a_i\right) L_0^{\sqrt{2^{m-1}N}}
\end{align*}
and so
\begin{align*}
\log P_N^{-1}
&= -\left(\sum_{i=1}^{\log{\sqrt{2^{m-1}N}}} \log a_i\right) + \sqrt{2^{m-1}N}\log L_0^{-1}.
\end{align*}

Note that $\log L_0^{-1}>0$, and for each $i$,
$0 \leq \log a_i \leq 1$. Thus,
\[
\log P_N^{-1} \leq \sqrt{2^{m-1}N}\log L_0^{-1}.
\]
Finally,
\begin{align*}
\log P_N^{-1}
&\geq -\left(\sum_{i=1}^{\log{\sqrt{2^{m-1}N}}} \right) + \sqrt{2^{m-1}N}\log L_0^{-1} \\
&= -\log{\sqrt{2^{m-1}N}} + \sqrt{2^{m-1}N}\log L_0^{-1} \\
&= \sqrt{2^{m-1}N} \left(\log L_0^{-1} -
\frac{\log{\sqrt{2^{m-1}N}}}{\sqrt{2^{m-1}N}}\right).
\end{align*}

For the case where $\log N-m$ is even, the proof is similar and it is omitted.

\section{Proof of Proposition 7}
Without loss of generality, we consider the upper half of $S$, denoted
by $S_\mathcal{U}$. As we shall see, the image of $S_\mathcal{U}$ is
exactly the reflection of $S_\mathcal{U}$ with respect to the line
$\beta=\alpha$ (denoted by $S_\mathcal{L}$).  We know that
$S_\mathcal{U}:=\{(\alpha,\beta)\in \mathcal{U}|\beta\leq\sqrt{\alpha}
\text{ and } \beta\geq1-(1-\alpha)^2\}$.

The image of $S_\mathcal{U}$ under $f$ can be calculated by
\[
(\alpha', \beta')=f(\alpha, \beta)=(1-(1-\alpha)^2, \beta^2),
\]
where $(\alpha, \beta)\in \mathcal{U}$.
The above relation is equivalent to
\[
(\alpha, \beta)=(1-\sqrt{1-\alpha'}, \sqrt{\beta'}).
\]
Therefore, we can calculate images of boundaries for $R_\mathcal{U}$ under $f$.

The image of the upper boundary $\beta \leq \sqrt{\alpha}$ is
\[
 \sqrt{\beta'} \leq \sqrt{1-\sqrt{1-\alpha'}},
\]
i.e.,
\[
\alpha' \geq 1-(1-\beta')^2,
\]
and that of  the lower boundary $\beta \geq 1-(1-\alpha)^2$ is
\[
\sqrt{\beta'} \geq 1-(1-(1-\sqrt{1-\alpha'}))^2,
\]
i.e.,
\[
\alpha' \leq \sqrt{\beta'}.
\]

The function $f$ is monotone. Hence, images of boundaries of
$S_\mathcal{U}$ are boundaries of $S_\mathcal{L}$. Notice that
boundaries of $R_\mathcal{L}$ are symmetric with those of
$R_\mathcal{U}$ about $\beta=\alpha$.  We conclude that $S$ is an
invariant region.

\ifCLASSOPTIONcaptionsoff
  \newpage
\fi

\bibliographystyle{IEEEbib}

\end{document}